\documentclass[preprint,12pt]{elsarticle}
\usepackage{amsmath, amssymb, amscd, amsthm, amsfonts}  
\usepackage{ upgreek }
\usepackage{graphicx} 
\usepackage{subcaption} 
\usepackage{hyperref} 
\usepackage{algorithm} 
\usepackage{algorithmicx}
\usepackage{algpseudocode} 
\usepackage{caption} 
\usepackage{float} 
\usepackage{indentfirst} 
\usepackage[T1]{fontenc} 
\usepackage[utf8]{inputenc} 
\usepackage{framed}
\usepackage{diagbox}
\usepackage{adjustbox}
\usepackage{bm}
\usepackage{booktabs}
\usepackage{multirow}
\usepackage{makecell}
\usepackage{setspace}
\usepackage{xcolor}

\newtheorem{theorem}{Theorem}
\newtheorem{lemma}{Lemma}
\newtheorem{definition}{Definition}
\newtheorem{remark}{Remark}

\journal{ }
\date{}

\begin{document}

\begin{frontmatter}

\title{Analysis of Maximum Threshold and Quantum Security for Fault-Tolerant Encoding and Decoding Scheme Base on Steane Code}

\author[1,2,3]{Qiqing Xia}
\author[4]{Huiqin Xie}
\author[1,2]{Li Yang  \corref{mycorrespondingauthor}}
\cortext[mycorrespondingauthor]{Corresponding author: yangli@iie.ac.cn}

\affiliation[1]{organization={Key Laboratory of Cyberspace Security Defense},
            city={Beijing},
            postcode={100093}, 
            country={China}}

\affiliation[2]{organization={ Institute of Information Engineering, Chinese Academy of Sciences},
            city={Beijing},
            postcode={100093}, 
            country={China}}
            
\affiliation[3]{organization={School of Cyber Security, University of Chinese Academy of Sciences},
            city={Beijing},
            postcode={100049}, 
            country={China}}

\affiliation[4]{organization={Beijing Electronic Science and Technology Institute},
            city={Beijing},
            postcode={100070}, 
            country={China}}

\begin{abstract}
Steane code is one of the most widely studied quantum error-correction codes, which is a natural choice for fault-tolerant quantum computation (FTQC). However, the original Steane code is not fault-tolerant because the CNOT gates in an encoded block may cause error propagation. In this paper, we first propose a fault-tolerant encoding and decoding scheme, which analyzes all possible errors caused by each quantum gate in an error-correction period. In this scheme, we combine the results of measuring redundant qubits with those of syndrome measurements to identify specific errors for different types of errors (X, Y, and Z-type errors). But due to the error propagation, there may be cases where different errors produce the same measurement results. Therefore, we introduce the "flag qubits" scheme (providing its usage conditions) to reduce error interference as much as possible, and we consider the errors caused by the introduced quantum gates, realizing the truly fault-tolerant Steane code. Afterwards, we provide the fault-tolerant scheme of the universal quantum gate set, including fault-tolerant preparation and verification of ancillary states. This is the first time that fault tolerance has been considered for every process of FTQC. Finally, We propose an algorithm for a more accurate estimation of thresholds and optimal error-correction period selection. Our simulation results based on this entire scheme demonstrate the effectiveness of this algorithm, satisfying the threshold theorem and the currently widely recognized threshold. We analyze the relationship among the maximum threshold, concatenated levels, and quantum logical depth, showing that quantum operations play a crucial role in increasing the threshold. This idea can be extended to other Calderbank-Shor-Steane (CSS) codes to improve the reliability of FTQC. Furthermore, we analyze the computational theoretical limits of quantum computers from the perspectives of attack and active defense based on our FTQC scheme, thereby assessing the security of a system.
\end{abstract}


\end{frontmatter}

\section{Introduction}\label{Introduction}
Quantum computation is a combination of quantum physics and computer science. It uses a quantum mechanical system as computational hardware, encodes data information with quantum states, performs computational tasks (transformation and evolution) according to the laws of quantum mechanics, and extracts computation results based on quantum measurement theory. On quantum computers, parallel computation can be performed by utilizing quantum phenomena such as quantum superposition and entanglement, enabling effective solutions to problems that are challenging for classical computers. According to Shor's algorithm \cite{shor1994algorithms,shor1994polynomial}, adversaries can use quantum computers to attack cryptosystems based on discrete logarithms or factorization problems in polynomial time, which causes a great threat to public key cryptosystems. Grover's algorithm \cite{grover1996fast,grover1997quantum} achieves a quadratic speedup compared to the exhaustive search in an unstructured database, which requires doubling the key length of symmetric cryptosystems to maintain classical security. However, since quantum computers are essentially physical systems, their specific implementation is constrained by many physical factors. Moreover, as quantum computation technology is still in its early stages, expanding it to quantum computers capable of producing effective computations will encounter significant difficulties.

One of the difficulties is decoherence \cite{landauer1995quantum, chuang1995quantum, unruh1995maintaining, landauer1996physical}. Quantum computation involves coherent quantum superposition states, which are often prone to decay, a phenomenon known as decoherence. One way to overcome decoherence is to measure the state of the quantum system through interactions between the environment and the quantum system \cite{zurek1991decoherence}. Another difficulty is the inability to realize quantum computation with perfect accuracy \cite{bernstein1993quantum, bennett1997strengths}. Errors in quantum gates can accumulate during the computation process, leading to inaccuracies beyond the tolerable threshold and eventually resulting in failure. The two difficulties are closely related. Decoherence can be represented by the inaccuracy of the quantum system state and the auxiliary quantum system interacting with it. Therefore, those methods to prevent decoherence can often be used to correct inaccuracy. Quantum error-correction codes \cite{gottesman2002introduction, shor1995scheme, steane1996error, steane1996multiple, calderbank1996good} can significantly reduce decoherence and inaccuracy in quantum data transmission and storage, advancing the development of quantum computation. For large-scale quantum computation, FTQC is the most practical candidate approach \cite{shor1996fault, preskill1998fault, gottesman1998theory, steane1999efficient}. In FTQC, error-correction codes add redundant qubits to data qubits for encoding. Each logical qubit is replaced by the encoded physical qubits, and the logic gates are replaced by the fault-tolerant gates. By periodically correcting errors, the accumulation of logical errors can be prevented.

Shor first constructed the quantum error correction scheme \cite{shor1995scheme}. This scheme uses nine physical qubits to encode one logical information qubit, detecting and correcting phase flip and phase flip errors on qubits, which lays the foundation for FTQC. Steane proposed another quantum error correction scheme \cite{steane1996error,steane1996multiple}, which uses seven physical qubits to encode one logical information qubit. Its relatively simple structure and efficient performance make it one of the most widely studied quantum error correction codes in the field of FTQC. Calderbank, Shor, and Steane proposed the system construction scheme for quantum error correction-CSS code \cite{steane1996error,calderbank1996good} based on the idea of classical linear block error-correction codes, which is widely used in the design of various quantum error-correction codes and the system provides more comprehensive protection. Based on the idea of CSS code, Bacon proposed the Bacon-Shor code \cite{bacon2006operator}, which uses symmetry to simplify the structure of error correction and improve the efficiency of the quantum error-correction code. Bravyi and Kitaev introduced the concept of quantum topological codes \cite{bravyi1998quantum}, placing physical qubits on a colored Latin lattice, where each stabilizer is only related to a few nearby qubits. Subsequently, Kitaev proposed a method for implementing FTQC by using topological quantum codes and anyons, where surface codes are an important class of topological quantum codes \cite{kitaev2003fault}. Surface codes can effectively detect and correct local errors, which provides a theoretical basis for large-scale quantum computation.

There are many forms of quantum error-correction codes, but the key to error-correction schemes is the fault-tolerant threshold in quantum theory. Concatenated codes \cite{knill1996concatenated} play an important role in determining thresholds because they can iteratively suppress inaccuracy by increasing the number of concatenated levels. In FTQC, the encoding and decoding circuits have a certain logical depth; the fault-tolerant logic gates in a quantum algorithm further increase the logical depth, especially considering that the physical realization of any single-qubit gate is approximated with arbitrary accuracy by several Hadamard gates and T gates, which will result in a large number of quantum operations on the same physical qubit. In addition, FTQC needs to be combined with concatenated codes for quantum error correction \cite{nielsen2010quantum,gottesman2010introduction}, which increases the logical depth acting on a physical qubit. Obviously, due to the constraints imposed by the physical properties of quantum computers, the error accumulation of multiple quantum operations may cause the inaccuracy to exceed the tolerated threshold, and quantum computation may not be executed reliably. However, the threshold theorem states that as long as the noise affecting computer hardware is less than a certain critical value, i.e., the accuracy threshold \cite{aharonov1997fault,aliferis2007accuracy,knill1998resilient,knill1996threshold}, quantum computation of any scale can be reliably executed. It is a natural choice based on Steane code in CSS code for fault-tolerance analysis because it is very small. There have been many studies on threshold analysis for Steane code \cite{preskill1998fault,nielsen2010quantum,zalka1996threshold,aliferis2005quantum}. Different thresholds depend on the assumptions and parameters of different FTQC schemes. Currently, the widely recognized threshold for Steane code is of the order $10^{-4}$ \cite{nielsen2010quantum}.

The original Steane code does not consider error propagation during the encoding and decoding processes (mostly assuming that encoding and decoding are perfect), therefore, in fact, these processes are not fault-tolerant, and thus the threshold analysis is incomplete. Due to the effect of the CNOT gates, the propagation of a single error may cause errors in multiple qubits, and only syndrome measurement through stabilizers alone cannot detect and correct multiple-qubit errors simultaneously. Recently, many new schemes have realized error-correction protocols based on "flag qubits" and identify multiple-qubit errors \cite{chao2018quantum,chamberland2018flag,tansuwannont2020flag,chao2018fault,chamberland2019fault,reichardt2020fault}, using the minimum number of auxiliary qubits to measure the stabilizers. In \cite{quan2022implementation}, a method based on the Steane stabilizers and "flag qubits" is provided to detect and correct error propagation caused by a certain CNOT gate in an encoded block, but it assumes that the first two CNOT gates and introduced auxiliary quantum gates are perfect.

\textbf{Our contributions} In this paper, we introduce this idea and propose a fault-tolerant encoding and decoding scheme based on Steane code, considering errors on each quantum operation. To our knowledge, no such specific analysis has been done for each quantum operation before. We also propose a more accurate algorithm to estimate the maximum threshold and obtain some conclusions through simulation. In more detail, our main contributions are as follows:
\begin{itemize}
    \item  We propose a scheme to implement fault-tolerant encoding and decoding based on Steane code. Steane states are employed for syndrome measurements, effectively reducing the number of the CNOTs during this process. Combined with the results of measuring redundant qubits during the decoding process, we can detect errors caused by each quantum gate in an error-correction period. However, there may be interference due to error propagation, i.e., different errors leading to the same measurement results. To address this, we introduce the "flag qubits" scheme and provide its usage conditions, aiming to minimize the interference as much as possible. We consider the error propagation for different types of errors in our scheme, including errors caused by the introduced auxiliary quantum gates with flag qubits. In addition, we provide the fault-tolerant scheme of the universal quantum gate set. This is the first time considering fault tolerance in all processes, including fault-tolerant encoding, fault-tolerant quantum gates, fault-tolerant decoding, as well as fault-tolerant preparation and verification of ancillary states. The fault-tolerant encoding and decoding scheme can be extended to other CSS codes, enhancing the reliability of FTQC.
    
    \item We propose a more accurate algorithm for estimating the threshold and selecting the optimal error-correction period. Combined with the permitted logical depth of the ion trap computer, the simulation results based on our entire FTQC scheme show that as the number of concatenated levels increases, the optimal selection for an error-correction period is to only execute one fault-tolerant quantum gate operation of the algorithm. Moreover, the maximum threshold we obtain is consistent with the widely recognized threshold \cite{nielsen2010quantum} currently. We observe that the threshold can be increased by increasing the number of concatenated levels and the logical depth of physical qubits in the auxiliary block before the fault-tolerant measurements. However, the limit value for increasing the threshold is the maximum threshold within the optimal error-correction period in the auxiliary block after fixing the number of concatenated levels. This algorithm can also be applied to other quantum error-correction codes.
    
    \item Most cryptographic systems currently are based on computational assumptions of mathematically hard problems \cite{portmann2022security}, such as factorization and discrete logarithm problems \cite{shor1994algorithms,shor1994polynomial}. Their security relies on computational hardness assumptions. Therefore, from the perspective of the underlying operations of quantum computers, we analyze the running time for performing an attack quantum algorithm based on our entire FTQC scheme, thereby assessing whether a cryptographic system can be broken within a meaningful time. We study the computational theoretical limits of quantum computers that can provide design guidelines under quantum computation environments, which lays the foundation for the idea of active defense.
\end{itemize}

\textbf{Outline} Section \ref{Preliminaries} introduces the relevant knowledge of logic gates and FTQC. Section \ref{fault-tolerant implementation based on Steane code} provides a scheme to implement fault-tolerant encoding and decoding based on Steane code. Section \ref{threshold analysis and quantum security} provides the fault-tolerant scheme for the universal quantum gate set and proposes a more accurate algorithm for estimating the threshold, followed by simulations and quantum security analysis. Section \ref{discussion} discusses some research issues that our work can further explore in the future. Section \ref{conclusion} summarizes the full paper and gives conclusions.

\section{Preliminaries}\label{Preliminaries}
In this section, we introduce the relevant knowledge of quantum computation and quantum information, including universal logic gates, fault-tolerant quantum computation, and concatenated codes.

\subsection{Logic Gates}
\begin{figure}[H]
    \centering
    \includegraphics[scale=0.85]{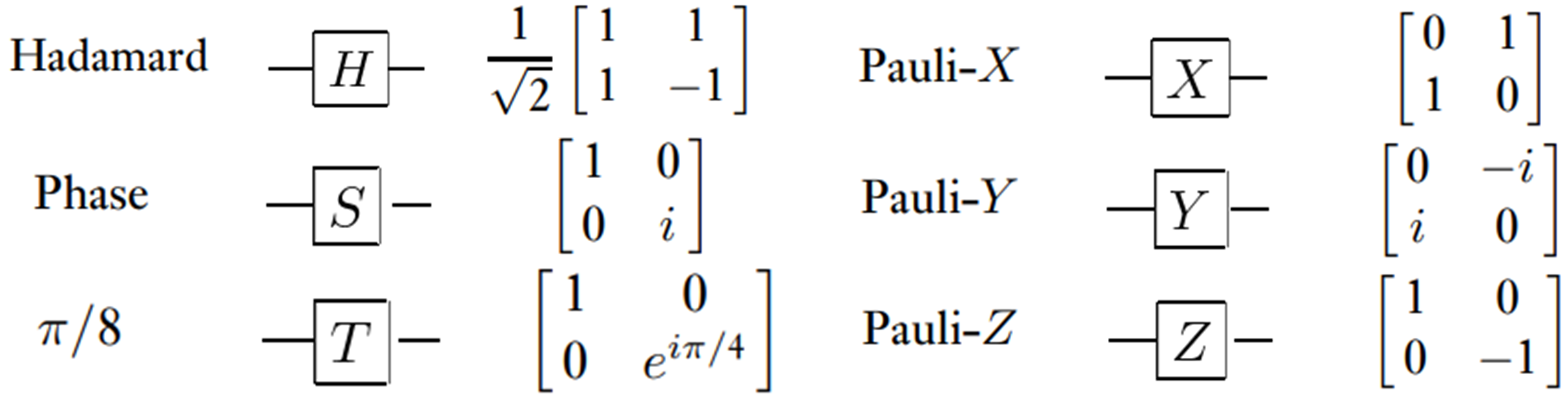}
    \caption{single-qubit quantum gates}
    \label{fig:sqqg}
\end{figure}

We briefly recall some single-qubit gates, including Hadamard gates (H gates), Pauli-X (Y, Z) gates, Phase gates (S gates), and $\frac{\pi}{8}$ gates (T gates). Their circuit symbols and matrix representations are shown in Figure \ref{fig:sqqg}.
\noindent Here, for the Pauli matrix (X, Y, Z), we have $Y=iXZ$.

\begin{figure}[H]
\centerline{\includegraphics[scale=1]{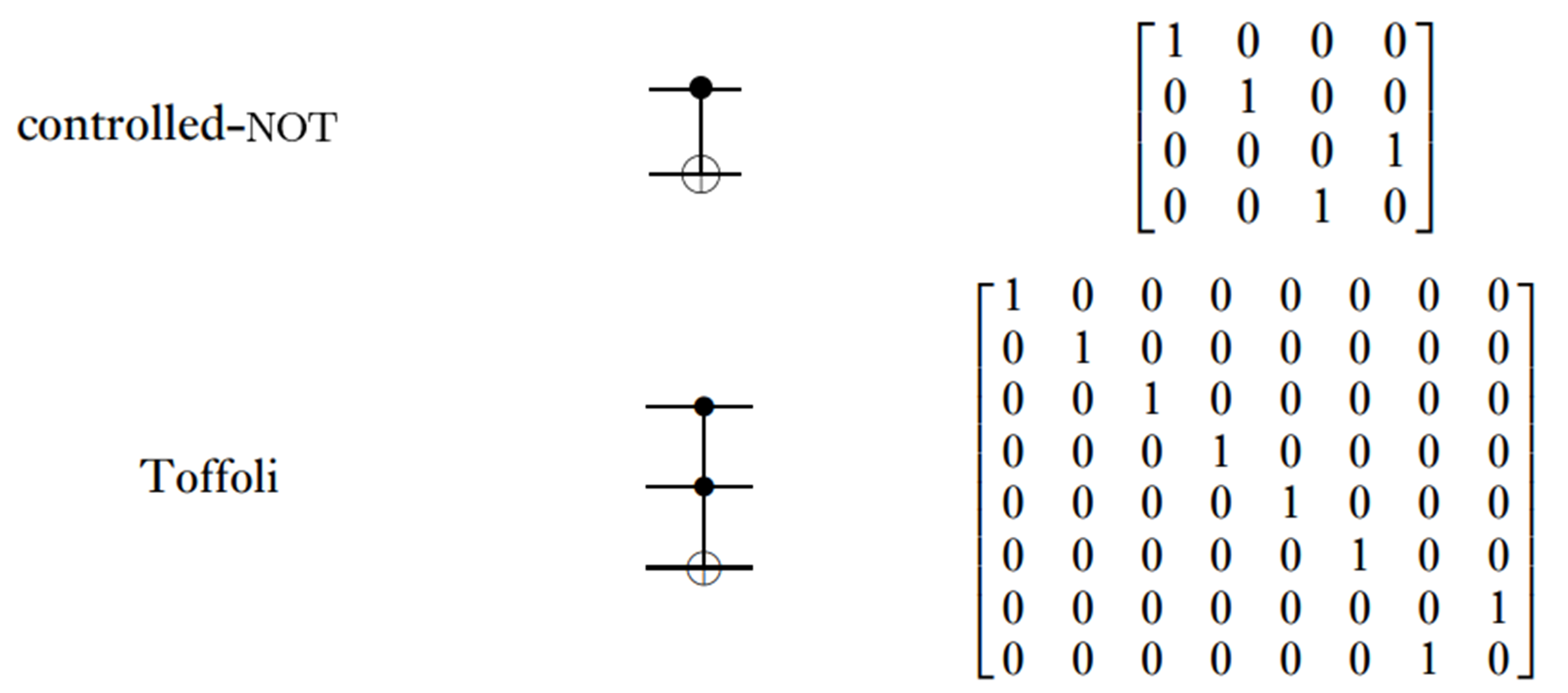}}  
  \caption{Multiple-qubit quantum gates}
   \label{fig:mqqg}
\end{figure}

Then, we recall two universal multi-qubit gates, namely controlled-NOT (CNOT) gates and Toffoli gates. Their circuit symbols and matrix representations are shown in Figure \ref{fig:mqqg}.

In quantum computation, single-qubit quantum gates and CNOT gates are universal, and their combination can realize any unitary operation. A single-qubit quantum gate can be approximated by H, S, and T gates with arbitrary accuracy. Therefore, \{H, S, T, CNOT\} can form a universal quantum gate set, capable of describing any quantum computation.

\begin{figure}[H]
    \centering
    \includegraphics[scale=0.3]{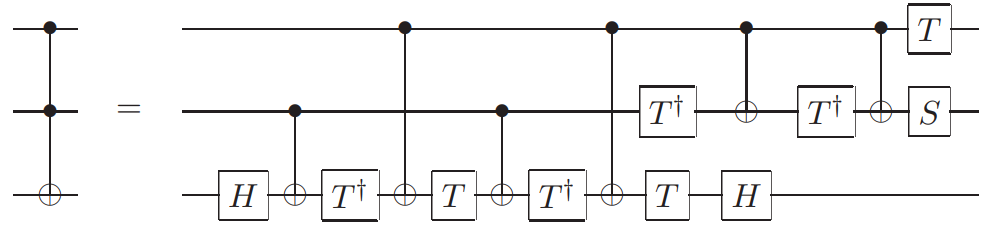}
    \caption{Decomposition of the Toffoli Gate}
    \label{fig:DOTTG}
\end{figure}

The Toffoli gate is a three-qubit quantum gate that can implement the "AND" operation in quantum computation and is regarded as a double-controlled CNOT gate. It can be decomposed into seven T gates, six CNOT gates, two H gates, and one S gate \cite{nielsen2010quantum}. The decomposed quantum circuit is shown in Figure \ref{fig:DOTTG}.

Typically, an $n$-controlled unitary transformation can be decomposed into $2n-2$ Toffoli gates after adding $n-1$ auxiliary qubits. The universal quantum gate set can also be replaced with \{H, S, CNOT, Toffoli\}. Therefore, Toffoli gates play a very important role in quantum computation as well.

\subsection{Fault-tolerant quantum computation}
Noise poses a significant threat to quantum systems, and therefore, efforts should be made to minimize its interference. FTQC utilizes quantum error-correction codes for encoding. Even if the information in the encoded message is affected by noise, there is sufficient redundancy to recover or decode the message. We first give some definitions about FTQC.

\begin{definition}\label{definition1}
A unitary operation is called a legal operation if it can map the code space onto itself. If this legal operation can be implemented by bit-wise operation, it is called a fault-tolerant operation.
\end{definition}

\begin{definition}\label{definition2}
In FTQC, the maximum number of operations in the encoded physical qubits for a fault-tolerant quantum gate to realize its logical function is called the logical depth of the fault-tolerant quantum gate.
\end{definition}

\begin{definition}\label{definition3}
If only one quantum operation fails in the encoded block, then the failure causes at most one error in each output block from the procedure. This operation process includes not only the quantum gate operation, but also the measurement with noise and the state preparation with noise. This property is called quantum fault tolerance.
\end{definition}

\begin{figure}[H]
\centerline{\includegraphics[scale=0.55]{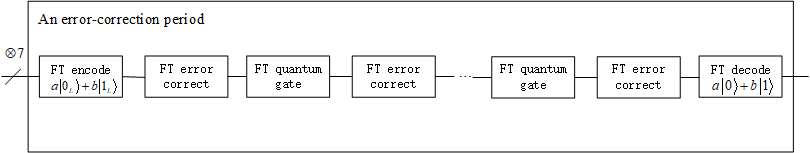}}  
  \caption{Quantum error-correction period circuit}
   \label{fig:qepc}
\end{figure}

To prevent the impact of noise, we apply Steane code for error correction. The quantum error correction process includes two steps: error syndrome measurement and recovery. Replacing a logical qubit with an encoded block and replacing quantum logic gates with fault-tolerant quantum gates, error correction operations are periodically performed to prevent the accumulation of inaccuracy on the encoded state. The error-correction period circuit is shown in Figure \ref{fig:qepc}.

\begin{figure}[H]
\centerline{\includegraphics[scale=0.7]{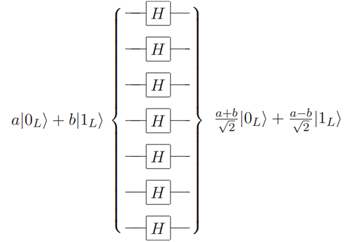}}  
  \caption{Transversal Hadamard gate based on Steane code}
   \label{fig:thg}
\end{figure}

According to Definitions \ref{definition1} and \ref{definition3}, quantum logic gates must maintain quantum fault tolerance, meaning the fault-tolerant quantum gate operation on an encoded state $a|0_L\rangle+b|1_L\rangle $ can be transversally implemented by qubit-wise. Taking Hadamard gate as an example, the transversal implementation based on Steane code is shown in Figure \ref{fig:thg}.
Namely, $\overline{H}=H^{\otimes 7}$. Similarly, S gates and CNOT gates can also be transversally implemented. As mentioned in section 2.1, T gates and Toffoli gates are elements of the universal quantum gate set. Although they cannot be directly transversally implemented, they can still be realized through fault-tolerant constructions (see section 4.1 for details).

\subsection{Concatenated code}
\begin{figure}[H]
\centerline{\includegraphics[scale=0.65]{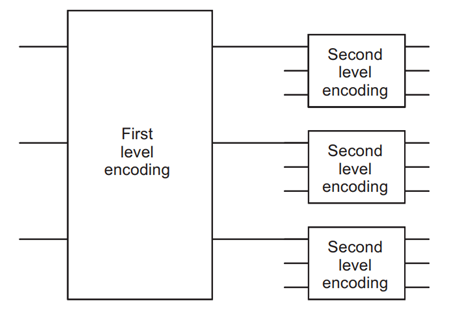}}  
  \caption{Concatenated code construction}
   \label{fig:ccc}
\end{figure}

To reduce the actual error probability, we can use the concatenated code construction. The idea is to iteratively apply the encoded circuit to simulate a logic circuit, and the construction is shown in Figure \ref{fig:ccc}.

Now consider the issues related to the threshold theorem. The threshold theorem is the core of FTQC, which shows that when the error probability (failure probability) of each fault-tolerant quantum gate (physical component) is less than a critical value, the arbitrarily long quantum computation can be performed reliably. First, we introduce a lemma \cite{nielsen2010quantum}:
\begin{lemma}\label{lemma1}
We denote $p$ as the failure probability for any component in a quantum circuit. Given a fault-tolerant quantum circuit, including encoding, fault-tolerant operations, syndrome measurement, and recovery, the probability that this circuit introduces two or more errors into the encoded block is $cp^2$, where $c$ is a constant.
\end{lemma}

According to Lemma $\ref{lemma1}$, if we concatenate $k$ levels, the failure probability for the top level is $(cp)^{2^k}/c$. Assuming we want to simulate a circuit containing polynomial $p(n)$ logic gates with accuracy $\epsilon$, each gate in the circuit must be simulated with an accuracy of $\epsilon/p(n)$. The concatenated levels satisfy:
\[\frac{(cp)^{2^k}}{c}\leq \frac{\epsilon}{p(n)}. \]
If $p<p_{th}\equiv 1/c$, such a $k$ can be found. This condition is called the threshold condition for quantum computation. The threshold theorem can be expressed as follows \cite{nielsen2010quantum}:

\begin{theorem}\label{theorem1}
A quantum circuit containing $p(n)$ logic gates can  be simulated with error probability at most $\epsilon$ using
\[\mathcal{O}(poly(\log(p(n)/\epsilon))p(n))\]
gates on hardware. Here, the failure probability of hardware components is at most $p$, assuming that $p$ meets the threshold condition $p<p_{th}$, and reasonable assumptions are made about the noise in the underlying hardware. 
\end{theorem}

\section{Fault-tolerant implementation based on Steane code}\label{fault-tolerant implementation based on Steane code}
In this section, we apply the Steane code to address noise in quantum systems. It is a [7, 1, 3] quantum error-correction code, which encodes one logical qubit into seven physical qubits and can simultaneously correct bit flip and phase flip errors on a physical qubit \cite{steane1996error,steane1996multiple}. From the perspective of the underlying physical components, the error probability on the control and target qubits of a physical CNOT gate is mutually independent. For each quantum operation, the non-zero error probability is very small. If errors on different quantum operations in the same block are completely uncorrelated with each other, then the probability of two errors occurring simultaneously is even smaller. Therefore, we can safely focus on the case where at most one error occurs on a quantum operation in each block.

We know that the original Steane code is not fault-tolerant, as there may be error propagation in an encoded block. Many current schemes make a fault-tolerant assumption about it (i.e., the original Steane code does not consider error propagation) and then recover data information through syndrome measurements. However, error recovery will not be perfect, as the recovery itself is a form of quantum computation and is prone to errors. Therefore, we must systematically consider the error possibility of each quantum operation resulting in recovery failure to ensure the overall process is fault-tolerant.

\subsection{Fault-tolerant encoding and decoding}
We propose a scheme to implement fault-tolerant encoding and decoding based on Steane code. Since errors may propagate when utilizing auxiliary quantum gates for FTQC, we not only need to consider all possible errors during the encoding and decoding processes, but also the errors caused by introduced auxiliary quantum gates.

\begin{figure}[H]
\centerline{\includegraphics[scale=0.5]{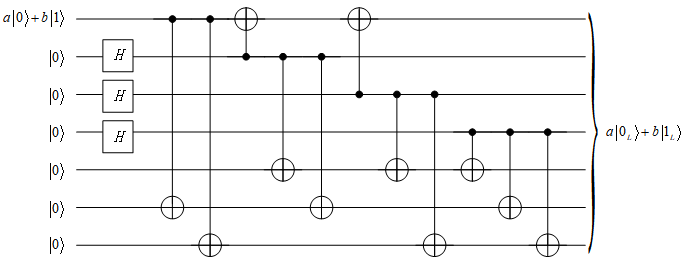}}
  \caption{Encoding circuit for Steane code}
   \label{fig:ecfsc}
\end{figure}

The original Steane code is closely related to the classical [7,4,3] Hamming code and can encode an unknown state using the circuit shown in Figure \ref{fig:ecfsc}.

After encoding, the logical $|0_L\rangle$ is the equally weighted superposition of all even weight codewords (with the even number of 1's) in the Hamming code, i.e.,
\begin{equation}
    \begin{aligned}
    |0_L\rangle&=\frac{1}{\sqrt{8}}\left(\sum_{\text{even}\ v\in Hanmming} |v\rangle \right)\\ &=\frac{1}{\sqrt{8}}\left(|0000000\rangle + |0001111\rangle + |0110011\rangle + |0111100\rangle \right.\\
    & \left. \quad \quad \quad + |1010101\rangle + |1011010\rangle + |1100110\rangle + |1101001\rangle \right),
    \end{aligned}
\end{equation}
and the logical $|1_L\rangle$ is the equally weighted superposition of all odd weight codewords (with the odd number of 1's) in the Hamming code, i.e.,
\begin{equation}
    \begin{aligned}
    |1_L\rangle&=\frac{1}{\sqrt{8}}\left(\sum_{\text{odd}\ v\in Hanmming} |v\rangle \right)\\ &=\frac{1}{\sqrt{8}}\left(|1111111\rangle + |1110000\rangle + |1001100\rangle + |1000011\rangle \right.\\
    & \left. \quad \quad \quad + |0101010\rangle + |0100101\rangle + |0011001\rangle + |0010110\rangle \right).
    \end{aligned}
\end{equation}

It is easy to understand the working principle of the encoder by using the alternative expression of the Hamming parity check matrix, 
\begin{equation}\label{eq3}
\begin{aligned}
H=[H_z|H_X]= 
\begin{bmatrix}
\begin{array}{c|c}
    1\quad0\quad1\quad0\quad1\quad0\quad1 & 0\quad0\quad0\quad0\quad0\quad0\quad0 \\
    0\quad1\quad1\quad0\quad0\quad1\quad1 & 0\quad0\quad0\quad0\quad0\quad0\quad0 \\
    0\quad0\quad0\quad1\quad1\quad1\quad1 & 0\quad0\quad0\quad0\quad0\quad0\quad0 \\
    0\quad0\quad0\quad0\quad0\quad0\quad0 & 1\quad0\quad1\quad0\quad1\quad0\quad1 \\
    0\quad0\quad0\quad0\quad0\quad0\quad0 & 0\quad1\quad1\quad0\quad0\quad1\quad1 \\
    0\quad0\quad0\quad0\quad0\quad0\quad0 & 0\quad0\quad0\quad1\quad1\quad1\quad1
\end{array}
\end{bmatrix},
\end{aligned}
\end{equation}
its stabilizer generators can be divided into Z-type stabilizers and X-type stabilizers, as follows:
\begin{equation}
    \begin{aligned}
&g_1^Z=Z_1I_2Z_3I_4Z_5I_6Z_7, g_2^Z=I_1Z_2Z_3I_4I_5Z_6Z_7, g_1^Z=I_1I_2I_3Z_4Z_5Z_6Z_7 \\
&g_1^X=X_1I_2X_3I_4X_5I_6X_7, g_2^X=I_1X_2X_3I_4I_5X_6X_7, g_1^X=I_1I_2I_3X_4X_5X_6X_7.
    \end{aligned}
\end{equation}

We consider three types of errors and treat them as mutually independent: bit-flip errors (X errors), phase-flip errors (Z errors), and bit-flip and phase-flip errors occurring simultaneously (Y errors, where Y=iXZ). When analyzing errors, it suffices to analyze X errors and Z errors separately, as Y errors are equivalent to the simultaneous occurrence of X and Z errors.

It can be seen from Figure \ref{fig:ecfsc} that there is error propagation in the encoded block of the original Steane code. We consider possible errors in all quantum gates, including errors that may occur in single-qubit quantum gates, CNOT gates, and introduced auxiliary quantum gates. We provide the possible error cases caused by Hadamard gates and CNOT gates, as follows:
\begin{enumerate}
    \item The auxiliary physical qubit (redundant qubits) used for encoding is $|0\rangle$, and after the Hadamard transform, $|+\rangle=\frac{1}{\sqrt{2}}(|0\rangle+|1\rangle)$ is obtained. If an X error on a Hadamard gate, the physical qubit will not be affected; if a Z error occurs, it will result in $|-\rangle=\frac{1}{\sqrt{2}}(|0\rangle- |1\rangle)$.
    
    \item If an X error occurs before the control qubit of a CNOT gate, the error will be propagated to the target qubit of the CNOT gate since $CNOT\cdot X\otimes I=X\otimes X \cdot CNOT$. Similarly, if a Z error occurs before the target qubit of a CNOT gate, the error will be propagated to the control qubit of the CNOT gate since $CNOT\cdot I\otimes Z=Z\otimes Z \cdot CNOT$.

    \item If an X error occurs before the target qubit of a CNOT gate or a Z error occurs before the control qubit of a CNOT gate, it will only affect the current physical qubit, and will not cause error propagation.
\end{enumerate}

The next step is the fault-tolerant syndrome measurement for the encoding. We first prepare an appropriate auxiliary state so that the measurement results will reveal the information about errors while not propagating errors into the encoded block. We have found a method that does not damage the quantum codewords when extracting syndromes, applying the auxiliary quantum state \cite{steane1997active}
\begin{equation}
    |Steane\rangle=\frac{1}{4}\sum_{v\in Hanmming} |v\rangle=H^{\otimes 7}|0_L\rangle=\frac{1}{\sqrt{2}}(|0_L\rangle+|1_L\rangle).
\end{equation}

\begin{figure}[H]
\centerline{\includegraphics[scale=0.7]{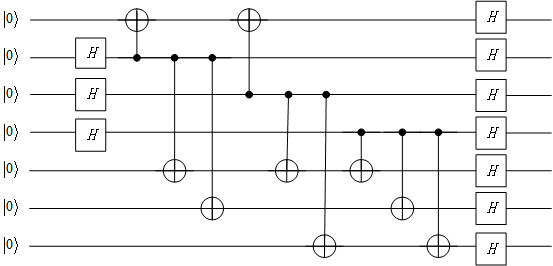}} 
  \caption{Construction of the Steane state}
   \label{fig:coss}
\end{figure}

If there are no errors, our measurement result is a random Hamming codeword and does not reveal the data information; if a physical qubit occurs an error, the check matrix $H$ in formula \ref{eq3} can be used to detect which qubit is in error. The same procedure is performed in a rotated basis to find the phase flip syndrome. Compared with Shor's method \cite{shor1996fault}, Steane's method only requires 14 auxiliary qubits and 14 CNOT gates, reducing the number of CNOT gates. The preparation process of $|Steane\rangle$ is shown in Figure \ref{fig:coss}.

Due to error propagation, a single error during the preparation of the Steane state may lead to two phase errors, thereby propagating into the data information. Therefore, multiple phase error tests must be performed on the Steane state to verify. If the tests fail, the state should be discarded, and a new Steane state must be constructed.

To verify the Steane state, we need to prepare two Steane states and perform the Hadamard transform, i.e., two encoded blocks $|0_L\rangle$, perform the qubit-wise XOR from the first block to the second block, and then measure the second block. We apply the classical Hamming check matrix to correct the bit-flip error and identify the measured block as $|0_L\rangle$ or $|1_L\rangle$. If the result is $|0_L\rangle$, the other block has passed the check. If the result is $|1_L\rangle$, then we flip the block to correct it.

However, this verification process is not yet reliable, as the measured block may be actually faulty. Therefore, we must repeat the verification step. If the measured block produces the same result twice, we can consider the check to be reliable. Otherwise, one of the measured blocks is faulty. If the error probability of each physical qubit in an encoded block is $\epsilon$, the failure probability of the checked block is $\epsilon^2$, which can be ignored. Through this verification process, we successfully constructed a Steane state.

Moreover, if a single error occurs during the syndrome measurement, it may introduce errors simultaneously in the data block and the ancillary block. Therefore, we must repeat the syndrome measurement \cite{preskill1998fault}. If we obtain the same result, we can safely accept the syndrome and recover, since the probability $\epsilon^2$ of the same (non-trivial) error syndrome twice is ignored. Otherwise, we can do nothing until the error is reliably detected in a later round of error correction.

\begin{figure}[H]
\centerline{\includegraphics[scale=0.55]{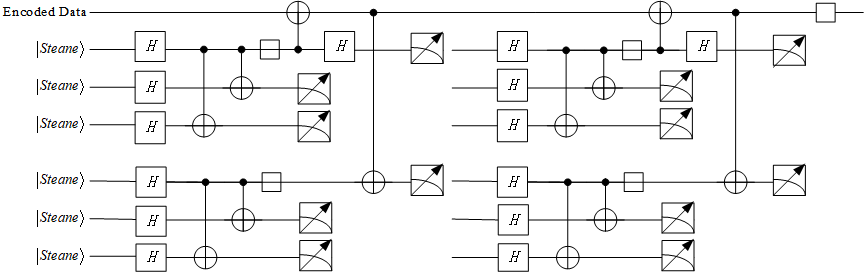}} 
  \caption{Error recovery circuit based on Steane states}
   \label{fig:9}
\end{figure}

Based on the analysis above, we provide the complete quantum circuit for fault-tolerant recovery, as shown in Figure \ref{fig:9}. Recovery will only fail when two independent errors of the same type occur, with probability $\epsilon^2$.
\noindent Here, both bit flip and phase flip are repeated twice, and the Steane state is also verified. "$\Box$" represents correction actions (conditioned on the measurement result).

\begin{figure}[H]
\centerline{\includegraphics[scale=0.4]{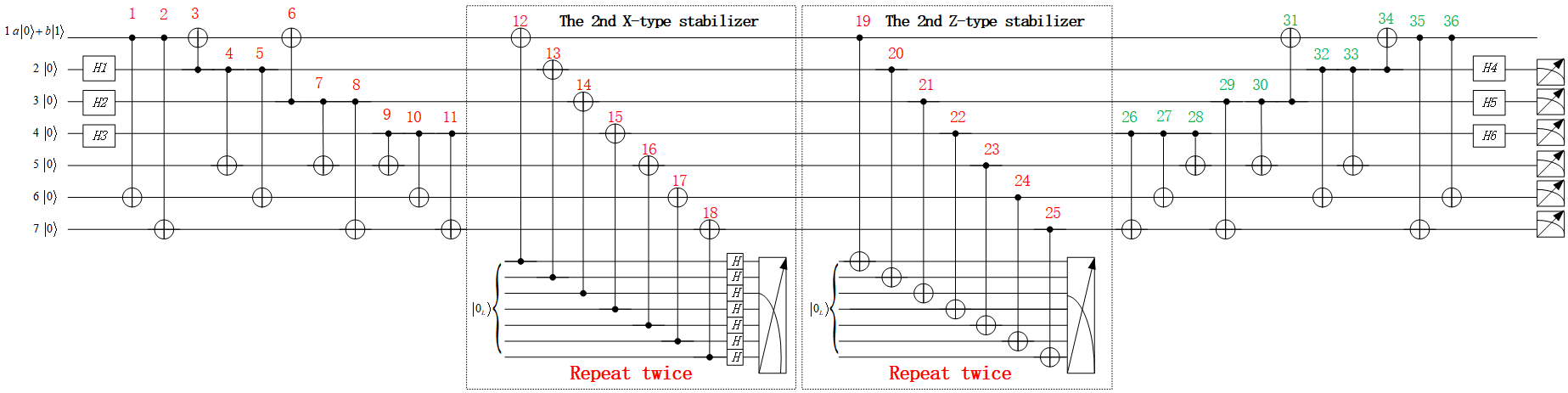}} 
  \caption{Fault-tolerant encoding and decoding based on Steane code}
   \label{fig:ftscbfq}
\end{figure}

Due to error propagation causing multiple physical qubits to be in error, the syndrome measurement alone is insufficient to detect which physical qubits are affected. Therefore, we need to combine the decoding process (the inverse process of encoding shown in Figure \ref{fig:ecfsc}), transfer errors to redundant qubits, and measure them. By combining the results of syndrome measurements with those of the redundant qubits, we can uniquely identify the error propagation caused by a specific quantum operation and correct it. A complete fault-tolerant encoding and decoding circuit based on Steane code is shown in Figure \ref{fig:ftscbfq} (the part related to the algorithmic fault-tolerant quantum gates and auxiliary quantum gates with flag qubits is not shown). Here, the part marked in green is the decoding CNOT operations, and the framed part is the fault-tolerant syndrome measurement that will not cause error propagation. The X-type stabilizers and Z-type stabilizers are repeated twice respectively. If two outcomes are different, it is considered that there is an error in the quantum operation during this process. Since $H^{\otimes 7}|Steane\rangle=|0_L\rangle$, we directly use $|0_L\rangle$ as the auxiliary state.

If an X(Z) error occurs on the control qubit (target qubit) of the $i$th CNOT gate, it is denoted as $X_i^{C(T)}(Z_i^{C(T)})$; if an X(Z) error occurs on the $i$th Hadamard gate $H_i$, it is denoted as $X_{H_i}(Z_{H_i})$; if an X(Z) error occurs on the $j$th physical qubit after decoding, it is denoted as $X_j(Z_j)$. When $X_i^{C(T)}(Z_i^{C(T)})$ results in an X(Z) error occurring on the jth physical qubit ultimately, it is denoted as $X_i^{C(T)}(Z_i^{C(T)}) \to X_j(Z_j)$. We denote the results of measuring redundant qubits (2nd-7th physical qubits in Figure \ref{fig:ftscbfq}) as $Meas_1, \cdots, Meas_6$, respectively. 

\subsection{Analysis of X-type error propagation}
First considering X-type errors, since an X error on $|+\rangle$ has no impact on the results, then quantum operations $X_{28}^{C}, X_{31}^{C}, X_{34}^{C}$ and $H_1-H_6$ do not require analysis and are not included in the logical depth. All possible outcomes are as follows:
\begin{center}
\fbox{\hspace{-1em}\begin{minipage}{1.15\textwidth}
\small
\setstretch{0.5}
\textbf{\hspace{1em}X-type Error}
\label{tb:1}
\begin{enumerate}
    \item When $g_1^Z, g_2^Z, g_3^Z$ is 000:
    \begin{itemize}
        \item \textcolor{brown}{$Meas_5, Meas_6, Meas_7$ is 000: $X_{36}^{C} \to X_1$ or No error;}
        \item \textcolor{brown}{$Meas_5, Meas_6, Meas_7$ is 010: $X_{24}^{C}=X_{27}^{T}=X_{32}^{T}=X_{36}^{T} \to X_6$ or $X_{35}^{C} \to X_1X_6$;}
        \item $Meas_5, Meas_6, Meas_7$ is 001: $X_{25}^{C}=X_{26}^{T}=X_{29}^{T}=X_{35}^{T} \to X_7$;
        \item $Meas_5, Meas_6, Meas_7$ is 011: $X_{19}^{C}=X_{30}^{C}=X_{31}^{T}=X_{33}^{C}=X_{34}^{T} \to X_1X_6X_7$;
        \item $Meas_5, Meas_6, Meas_7$ is 100: $X_{23}^{C}=X_{27}^{C}=X_{28}^{T}=X_{30}^{T}=X_{33}^{T} \to X_5$;
        \item \textcolor{cyan}{$Meas_5, Meas_6, Meas_7$ is 110: $X_{21}^{C} \to X_1X_5X_6$ or $X_{26}^{C} \to X_5X_6$;}
        \item $Meas_5, Meas_6, Meas_7$ is 101: $X_{20}^{C} \to X_1X_5X_7$;
        \item \textcolor{cyan}{$Meas_5, Meas_6, Meas_7$ is 111: $X_{22}^{C} \to X_5X_6X_7$ or $X_{29}^{C}=X_{32}^{C} \to X_1X_5X_6X_7$;}
    \end{itemize}

    \item When $g_1^Z, g_2^Z, g_3^Z$ is 100:
    \begin{itemize}
        \item $Meas_5, Meas_6, Meas_7$ is 011: $X_{2}^{C}=X_{3}^{T}=X_{6}^{T}=X_{3}^{C}=X_{6}^{C}=X_{12}^{T} \to X_1X_6X_7$;
    \end{itemize}

    \item When $g_1^Z, g_2^Z, g_3^Z$ is 010:
    \begin{itemize}
        \item $Meas_5, Meas_6, Meas_7$ is 101: $X_{5}^{C}=X_{13}^{T} \to X_1X_5X_7$;
    \end{itemize}

    \item When $g_1^Z, g_2^Z, g_3^Z$ is 110:
    \begin{itemize}
        \item \textcolor{cyan}{$Meas_5, Meas_6, Meas_7$ is 110: $X_{8}^{C}=X_{14}^{T} \to X_1X_5X_6$ or $X_{10}^{C} \to X_5X_6$;}
    \end{itemize}

    \item When $g_1^Z, g_2^Z, g_3^Z$ is 001:
    \begin{itemize}
        \item \textcolor{cyan}{$Meas_5, Meas_6, Meas_7$ is 111: $X_{4}^{C}=X_{7}^{C} \to X_1X_5X_6X_7$ or $X_{11}^{C}=X_{15}^{T} \to X_5X_6X_7$;}
    \end{itemize}

    \item When $g_1^Z, g_2^Z, g_3^Z$ is 101:
    \begin{itemize}
        \item $Meas_5, Meas_6, Meas_7$ is 100: $X_{4}^{T}=X_{7}^{T}=X_{9}^{C}=X_{9}^{T}=X_{16}^{T} \to X_5$;
    \end{itemize}

    \item When $g_1^Z, g_2^Z, g_3^Z$ is 011:
    \begin{itemize}
        \item \textcolor{brown}{$Meas_5, Meas_6, Meas_7$ is 010: $X_{1}^{C} \to X_1X_6$ or $X_{1}^{T}=X_{5}^{T}=X_{10}^{T}=X_{17}^{T} \to X_6$;}
    \end{itemize}

    \item When $g_1^Z, g_2^Z, g_3^Z$ is 111:
    \begin{itemize}
        \item $Meas_5, Meas_6, Meas_7$ is 001: $X_{2}^{T}=X_{8}^{T}=X_{11}^{T}=X_{18}^{T} \to X_7$.
    \end{itemize}
\end{enumerate}
\end{minipage}}
\end{center}

\begin{figure}[H]
\centerline{\includegraphics[scale=0.85]{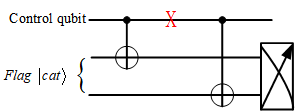}}
  \caption{Flag scheme to detect the X error on the control qubit}
   \label{fig:Flag-X}
\end{figure}

Theoretically, each unit should correspond to a unique error case. In \textbf{X-type Error}, the brown texts indicate errors that cannot be distinguished due to the same measurement result, while the blue texts indicate errors with the same measurement results that can be distinguished by the introduced ”flag qubits“ scheme. Additionally, we also need to avoid the potential errors introduced by auxiliary quantum gates with flag qubits. The scheme used for detecting X-type errors on the control qubits is shown in Figure \ref{fig:Flag-X}.

We use the two-qubit cat state as a Flag. When an X error occurs between the control qubits of two CNOT gates, the measurement result is 01 or 10, otherwise it is 00 or 11. This scheme avoids the propagation of Z-type errors on the target qubits. The prerequisites for using this scheme are as follows:
\begin{enumerate}
    \item The Flag measurement result of the first CNOT control qubit with an X error is the same as that of the first or second CNOT target qubit with an X error in Figure \ref{fig:Flag-X}. Therefore, the measurement results in \textbf{X-type Error} must be different to distinguish between the two cases, otherwise the scheme fails (unless both these cases have no impact on the results);
    \item An X error on the control qubit of the second CNOT cannot interfere with the existing measurement results in \textbf{X-type Error}, meaning it cannot result in different cases for the same measurement results;
    \item A Z error on the control qubits of the two CNOTs cannot interfere with the detection of \textbf{Z-type Error}.
\end{enumerate}

We use the Flag scheme in Figure \ref{fig:Flag-X} to detect $X_{4}^{C}$. Two CNOT gates denoted as CN1 and CN2 are introduced on the left side of the control qubit of the CNOT gate labeled as 3 and on the right side of the control qubit of the CNOT gate labeled as 5, respectively. The measurement result of an X error on the control qubit of CN1 is identical to that of an X error on the target qubit of CN1 or CN2, affecting the physical qubits $X_{1}X_{3}X_{5}X_{7}$. This is equivalent to No error and does not affect the result. The measurement result of an X error on the control qubit of CN2 is the same as that of $X_{13}^{T}$ and does not cause any interference with the existing measurement result (see \textbf{X-type Error}). Similarly, on both sides of the control qubits of the CNOT gates labeled as 6 and 8, as well as the CNOT gates labeled as 9 and 11, four CNOT gates are introduced, denoted as CN3 and CN4, CN5 and CN6, respectively. This allows for the detection of $X_{7}^{C}$, $X_{8}^{C}$ and $X_{10}^{C}$, $X_{11}^{C}$. 

Two CNOT gates denoted as CN7 and CN8 are introduced on the left and right sides of the control qubits of the CNOT gates labeled as 22 and 28 for detecting $X_{22}^{C}$ and $X_{26}^{C}$. If an X error occurs on the control qubit of CN7, resulting in different measurement results for the Z-type stabilizer repeated twice. If an X error occurs on the target qubits of CN7 or CN8, both $g_1^Z, g_2^Z, g_3^Z$ and $Meas_5, Meas_6, Meas_7$ are 000. This allows distinguishing between these two different cases. If an X error occurs on the control qubit of CN8, it does not interfere with the existing measurement results.  Therefore, combined with the above analysis, CN1-CN8 all meet the first two condition for using the Flag scheme in Figure \ref{fig:Flag-X}, and the blue texts indicate that different X-type error cases can be distinguished for same measurement results.

\begin{remark}\label{rm1}
The X error on the control qubit of CN8 replaces the original $X_{28}^{C}$ in \textbf{X-type Error}. Since the original $X_{28}^{C}$ has no impact on the results, the X error on the control qubit of CN8 also has no impact on the results. Therefore, the control qubit of CN8 does not need to be considered in the quantum logical depth of X-type errors.
\end{remark}

There are some error cases that cannot be judged due to the inherent structure of the Steane code. For example, $X_{36}^{C}$ is an X error in the last quantum operation of the decoding process and cannot be detected. Even if an auxiliary quantum gate is introduced by using the Flag scheme, it is inevitable that an error in the last auxiliary quantum gate operation causes the data qubit error. $X_{1}^{C}$, $X_{35}^{C}$, and $X_{24}^{C}$ cannot be detected by using the Flag scheme in Figure \ref{fig:Flag-X} because they do not satisfy the conditions for detecting X errors and may interfere the original measurement results, which is unavoidable. Due to the existence of different errors yielding the same measurement results, minimizing the assumption of perfect quantum operations is crucial for more accurate estimation of thresholds in subsequent analyses. Therefore, we need to minimize the number of assumed perfect quantum operations. The assumed perfect quantum operations include $X_{1}^{C}$, $X_{35}^{C}$, and $X_{36}^{C}$ in \textbf{X-type Error}.

\subsection{Analysis of Z-type error propagation}
Then considering Z-type errors, since a Z error on $|0\rangle$ has no impact on the results, then quantum operations $Z_{33}^{T}, Z_{35}^{T}, Z_{36}^{T}$ do not require analysis and are not included in the logical depth. All possible outcomes are as follows:
\begin{center}
\fbox{\begin{minipage}{1.25\textwidth}
\small
\setstretch{0.5}
\textbf{Z-type Error}
\label{tb:2}
\begin{enumerate}
    \item When $g_1^X, g_2^X, g_3^X$ is 000 :
    \begin{itemize}
        \item \textcolor{brown}{$Meas_1, Meas_2, Meas_3$ is 000: $Z_{1}^{C}=Z_{2}^{C}=Z_{1}^{T}=Z_{2}^{T}=Z_{29}^{T}=Z_{32}^{T}=Z_{34}^{T}=Z_{35}^{C}=Z_{36}^{C} \to Z_1$ or No error;}
        \item \textcolor{cyan}{$Meas_1, Meas_2, Meas_3$ is 010: $Z_{14}^{T}=Z_{21}^{C}=Z_{29}^{C}=Z_{30}^{C}=Z_{31}^{C}=X_{H_5} \to Z_3$ or $Z_{26}^{T} \to Z_1Z_3$;}
        \item $Meas_1, Meas_2, Meas_3$ is 001: $Z_{15}^{T}=Z_{22}^{C}=Z_{26}^{C}=Z_{27}^{C}=Z_{28}^{C}=X_{H_6} \to Z_4$;
        \item $Meas_1, Meas_2, Meas_3$ is 011: $Z_{18}^{T}=Z_{25}^{C} \to Z_1Z_3Z_4$;
        \item \textcolor{brown}{$Meas_1, Meas_2, Meas_3$ is 100: $Z_{13}^{T}=Z_{20}^{C}=Z_{30}^{T}=Z_{32}^{C}=Z_{33}^{C}=Z_{34}^{C}=X_{H_4} \to Z_2$ or $Z_{27}^{T}=Z_{31}^{T} \to Z_1Z_2$;}
        \item \textcolor{cyan}{$Meas_1, Meas_2, Meas_3$ is 110: $Z_{12}^{T}=Z_{19}^{C} \to Z_1Z_2Z_3$ or $Z_{28}^{T} \to Z_2Z_3$;}
        \item $Meas_1, Meas_2, Meas_3$ is 101: $Z_{17}^{T}=Z_{24}^{C} \to Z_1Z_2Z_4$;
        \item $Meas_1, Meas_2, Meas_3$ is 111: $Z_{16}^{T}=Z_{23}^{C} \to Z_2Z_3Z_4$;
    \end{itemize}

    \item When $g_1^X, g_2^X, g_3^X$ is 100:
    \begin{itemize}
        \item \textcolor{cyan}{$Meas_1, Meas_2, Meas_3$ is 110: $Z_{6}^{T} \to Z_1Z_2Z_3$ or $Z_{7}^{T} \to Z_2Z_3$;}
    \end{itemize}

    \item When $g_1^X, g_2^X, g_3^X$ is 010:
    \begin{itemize}
        \item \textcolor{brown}{$Meas_1, Meas_2, Meas_3$ is 100: $Z_{3}^{T}=Z_{5}^{T} \to Z_1Z_2$ or $Z_{H_1}=Z_{3}^{C}=Z_{4}^{C}=Z_{5}^{C}=Z_{4}^{T} \to Z_2$;}
    \end{itemize}

    \item When $g_1^X, g_2^X, g_3^X$ is 110:
    \begin{itemize}
        \item \textcolor{brown}{$Meas_1, Meas_2, Meas_3$ is 010: $Z_{H_2}=Z_{6}^{C}=Z_{7}^{C}=Z_{8}^{C} \to Z_3$ or $Z_{8}^{T} \to Z_1Z_3$;}
    \end{itemize}

    \item When $g_1^X, g_2^X, g_3^X$ is 001:
    \begin{itemize}
        \item $Meas_1, Meas_2, Meas_3$ is 101: $Z_{H_3}=Z_{9}^{C}=Z_{10}^{C}=Z_{11}^{C} \to Z_4$;
    \end{itemize}

    \item When $g_1^X, g_2^X, g_3^X$ is 101:
    \begin{itemize}
        \item $Meas_1, Meas_2, Meas_3$ is 111: $Z_{9}^{T} \to Z_2Z_3Z_4$;
    \end{itemize}

    \item When $g_1^X, g_2^X, g_3^X$ is 011:
    \begin{itemize}
        \item $Meas_1, Meas_2, Meas_3$ is 101:  $Z_{10}^{T} \to Z_1Z_2Z_4$;
    \end{itemize}

    \item When $g_1^X, g_2^X, g_3^X$ is 111:
    \begin{itemize}
        \item $Meas_1, Meas_2, Meas_3$ is 011:  $Z_{11}^{T} \to Z_1Z_3Z_4$.
    \end{itemize}
\end{enumerate}
\end{minipage}}
\end{center}

\begin{figure}[H]
\centerline{\includegraphics[scale=0.85]{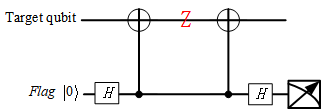}}
  \caption{Flag qubit to detect the Z error on the control qubit}
   \label{fig:Flag-Z}
\end{figure}

Similarly, we introduce the "flag qubits" scheme to distinguish different error cases with the same measurement results in \textbf{Z-type Error}. The scheme used for detecting Z-type errors on the target qubits is shown in Figure \ref{fig:Flag-Z}.

We use the zero quantum state as a Flag. When a Z error occurs between the target qubits of two CNOT gates, the measurement result is 1; otherwise, it is 0. This scheme does not cause the propagation of X-type errors on the control qubits. The prerequisites for using this scheme are as follows:
\begin{enumerate}
    \item The Flag measurement result of the first CNOT target qubit with a Z error is the same as that of the first or second CNOT control qubit with a Z error in Figure \ref{fig:Flag-Z}. Therefore, the measurement result in \textbf{Z-type Error} must be different to distinguish between the two cases, otherwise the scheme fails (unless both these cases have no impact on the results);
    \item A Z error on the target qubit of the second CNOT cannot interfere with the existing measurement results in \textbf{Z-type Error};
    \item An X error on the target qubits of the two CNOTs cannot interfere with the detection of \textbf{X-type Error}.
\end{enumerate}

We use the Flag scheme in Figure \ref{fig:Flag-Z} to detect $Z_{7}^{T}$. Two CNOT gates denoted as CN9 and CN10 are introduced on the left side of the target qubit of the CNOT gate labeled as 4 and on the right side of the target qubit of the CNOT gate labeled as 16, respectively. The measurement result of a Z error on the target qubit of CN9 is identical to that of a Z error on the control qubit of CN9 or CN10, affecting the physical qubits $X_{2}X_{3}X_{4}X_{5}$. This is equivalent to No error and does not affect the judgment. The measurement result of a Z error on the target qubit of CN10 is the same as that of $Z_{23}^{C}$ and does not cause any interference (see \textbf{Z-type Error}).

Two CNOT gates denoted as CN11 and CN12 are introduced on the left and right sides of the target qubits of the CNOT gates labeled as 26 and 35 for detecting $Z_{26}^{T}$ and $Z_{29}^{T}$. If a Z error occurs on the target qubit of CN11, resulting in $g_1^X,g_2^X,g_3^X$ being 000, and $Meas_1,Meas_2,Meas_3$ being 011. If a Z error occurs on the control qubits of CN11 or CN12, both $g_1^X,g_2^X,g_3^X$ and $Meas_1,Meas_2,Meas_3$ are 000. This allows distinguishing between these two different cases. If a Z error occurs on the target qubit of CN12, it does not interfere with the measurement results. Similarly, on the left and right sides of the target qubits of the CNOT gates labeled as 27 and 36, as well as the CNOT gates labeled as 28 and 33, four CNOT gates are introduced, denoted as CN13 and CN14, CN15 and CN16, respectively. This allows for detecting $Z_{27}^{T}$, $Z_{32}^{T}$ and $Z_{28}^{T}$
, $Z_{30}^{T}$. Therefore,  CN9-CN16 all meet the first two condition for using the Flag scheme in Figure \ref{fig:Flag-Z}, and the blue texts indicate that different Z-type error cases can be distinguished for the same measurement results.

\begin{remark}
The Z error on the target qubit of CN12 replaces the original $Z_{35}^{T}$ in \textbf{Z-type Error}. Since the original $Z_{35}^{T}$ has no impact on the results, the Z-type error on the target qubit of CN12 also has no impact on the results. Therefore, the target qubit of CN12 does not need to be considered in the quantum logical depth of Z-type errors. Similarly, the target qubits of CN14 and CN16 do not need to be considered in the quantum logical depth of Z-type errors.
\end{remark}

Firstly, Z errors on the control qubits cannot be detected by using the Flag scheme in Figure \ref{fig:Flag-Z}. Secondly, Z errors on the target qubits in the brown texts cannot be distinguished because they do not satisfy the conditions for detecting Z errors using the Flag scheme. Likewise, to minimize the assumptions of perfect quantum operations, the assumed perfect quantum operations include $Z_{1}^{C(T)}$, $Z_{2}^{C(T)}$, $Z_ {3}^{T}$, $Z_{5}^{T}$, $Z_{8}^{T}$, $Z_{31}^{T}$, $Z_{34}^{T}$, $Z_{35}^{C}$, $Z_{36}^{C}$ in \textbf{Z-type Error}.

To further examine whether other types of errors on the introduced auxiliary quantum gates with flag qubits would impact the encoding and decoding processes, we analyze the measurement results of Z-type errors on CN1-CN8 and X-type errors on CN9-CN16:
\begin{enumerate}
    \item First, we analyze the impact of Z-type errors on CN1-CN8. Since the quantum circuit we designed in Figure \ref{fig:Flag-X} does not propagate errors when the target qubits occur Z errors, we only need to consider the impact of Z-type errors on the control qubits. The measurement result of a Z-type error on the control qubit of CN1 or CN2 is the same as that of $Z_{H1}=Z_{3}^{C}=Z_{4}^{C}=Z_{5}^{C}=Z_{4}^{T}$ (see \textbf{Z-type Error}). Since we have assumed that $Z_{3}^{T}$ and $Z_{5}^{T}$ are perfect, the Z-type error on the control qubit of CN1 or CN2 can be detected. Similarly, the Z-type errors on the control qubits of CN3 or CN4 can also be detected. The measurement result of a Z-type error on the control qubits of CN5 or CN6 is the same as that of $Z_{H3}=Z_{9}^{C}=Z_{10}^{C}=Z_{11}^{C}$ (see \textbf{Z-type Error}), and there is no interference, so they can also be detected. Similarly, a Z-type error on the control qubits of CN7 and CN8 can be detected. Therefore, CN1-CN8 all meet the third condition for using the Flag scheme in Figure \ref{fig:Flag-X}.
    
    \item Next, we analyze the impact of X-type errors on CN9-CN16. The control qubits in Figure \ref{fig:Flag-Z} do not cause the propagation of X-type errors, so we only need to consider the impact of X-type errors on the target qubits. The measurement result of an X-type error on the target qubits of CN13 or CN14 is the same as that of $X_{24}^{C}=X_{27}^{T}=X_{32}^{T}=X_{36}^{T}$ (see \textbf{X-type Error}), and since we have assumed that $X_{35}^{C}$ is perfect, the X-type errors on the target qubits of CN13 and CN14 can be detected. We find that the X-type errors on the target qubits of other CN9-CN12, CN15, and CN16 do not interfere with the results in \textbf{X-type Error}, i.e., their errors do not appear in the brown texts of \textbf{X-type Error}.  Therefore, CN9-CN16 all meet the third condition for using the Flag scheme in Figure \ref{fig:Flag-Z}.
\end{enumerate}

To sum up, we assume that all quantum operations that cause X-type and Z-type errors with indistinguishable measurement results are perfect, and try to minimize the assumptions of perfect quantum quantum operations as much as possible, so that we can greatly demonstrate the real error correction effect based on Steane code. The Flag scheme reduces error interference, and a total of 16 CNOT gates have been introduced. We list the quantum operations that require perfect assumptions for X-type errors, Z-type errors, and Y-type errors:
\begin{itemize}
    \item For X-type errors, a total of 8 CNOTs have been introduced, we need to assume perfect quantum operations: $X_{1}^{C}, X_{35}^{C}$, and $X_{36}^{C}$. Therefore,  excluding perfect operations and operations that do not affect the results, the quantum logical depth of the first to seventh physical qubits in the encoded block during the encoding and decoding processes is 9, 11, 11, 12, 14, 12, and 12, respectively.
    
    \item For Z-type errors, a total of 8 CNOTs have been introduced, we need to assume perfect quantum operations: $Z_{1}^{C(T)}$, $Z_{2}^{C(T)}$, $Z_{3}^{T}$, $Z_{5}^{T}$, $Z_{8}^{T}$, $Z_{31}^{T}$, $Z_{34}^{T}$, $Z_{35}^{C}$, $Z_{36}^{C}$. Therefore, excluding perfect operations and operations that do not affect the results, the quantum logical depth of the first to seventh physical qubits in the encoded block during the encoding and decoding processes is 5, 14, 14, 16, 12, 8, and 8, respectively.
    
    \item For Y-type errors, they are equivalent to the simultaneous occurrence of X-type and Z-type errors, we need to assume perfect quantum operations: $Y_{1}^{C(T)}$, $Y_{2}^{C(T)}$, $Y_{3}^{T}$, $Y_{5}^{T}$, $Y_{8}^{T}$, $Y_{31}^{T}$, $Y_{34}^{T}$, $Y_{35} ^{C}$, $Y_{36}^{C}$. Excluding perfect operations, all quantum logical depth needs to be taken into account. The quantum logical depth of the first to seventh physical qubits in the encoded block during the encoding and decoding processes is 5, 14, 14, 16, 14, 10, and 10, respectively. 
\end{itemize}

\begin{remark}\label{rm3}
For the encoded $|0_L\rangle$ as an auxiliary block, the fault-tolerant encoding and decoding processes in Figure \ref{fig:ftscbfq} do not include the CNOT gates labeled as 1, 2, 35, 36. If we analyze the X-type errors in the auxiliary block $|0_L\rangle$, then all errors on every quantum operation can be detected without making perfect assumptions when combined with this Flag scheme in Figure \ref{fig:Flag-X}. Moreover, it is not necessary to analyze the Z-type errors, since $Z|0\rangle=|0\rangle$ after decoding.
\end{remark}

Combined with Remark \ref{rm1} and \ref{rm3}, we only need to consider the X-type errors and Y-type errors and their quantum logic depth is the same. Excluding the operations that do not affect the results, the quantum logical depth of the first to seventh physical qubits in the auxiliary block $|0_L\rangle$ during the encoding and decoding processes is all 8, 11, 11, 12, 10, 8, and 8, respectively.

\section{Threshold analysis and Quantum security}\label{threshold analysis and quantum security}
\subsection{Algorithm for estimating threshold}
In FTQC, we replace the logical gates in the algorithm with fault-tolerant quantum gates. To prevent the accumulation of errors, error correction must be performed periodically. Combined with section 2.3, it is believed that the optimal selection to obtain the maximum threshold is to perform an error correction process after each fault-tolerant quantum gate operation. In order to analyze the optimal error-correction period, we provide fault-tolerant implementation schemes for those universal quantum gates.

\begin{remark}\label{rm4}
As can be seen from section 2.1, the fault-tolerant quantum gates \{H, CNOT, S, Toffoli\} and \{H, CNOT, S, T\} are both universal quantum gate sets in FTQC. \{H,CNOT,S\} can be directly implemented transversally, and the logical depth of the fault-tolerant quantum gates is 1; \{T,Toffoli\} must implement fault tolerance through fault-tolerant measurement. After applying fault-tolerant T or Toffoli gate, the original encoded block needs to be measured, the logical operation is transferred to the auxiliary block and its result is stored in the auxiliary block. At this time, the logical depth of the fault-tolerant quantum gate in the original encoded block is 1, and the logical depth in the auxiliary block is greater than 1.
\end{remark}

\begin{figure}[H]
\centerline{\includegraphics[scale=0.85]{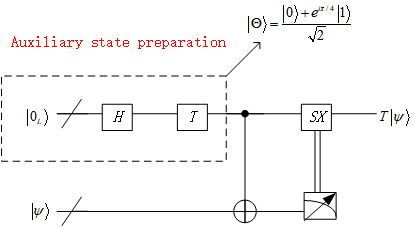}}
  \caption{Construction of fault tolerant T gate}
   \label{fig:cofttg}
\end{figure}

Next, we provide the fault-tolerant implementation schemes of \{T,Toffoli\} and explain the above analysis. We first analyze the logical depth of the fault-tolerant T gate in an encoded block and auxiliary block. Its fault-tolerant implementation scheme is shown in Figure \ref{fig:cofttg}.

Here, $|\Theta \rangle$ is the auxiliary quantum state, which is a +1 eigenstate of the single-qubit operator $e^{-i\pi/4}SX$. We still need the fault-tolerant measurements to prepare this auxiliary state. If the measurement result is +1, it can be considered to have been prepared correctly; if it is -1, a fault-tolerant $Z$ operation is applied to change the state. This process requires repeating the measurement twice. If the results are the same, we can ensure that the auxiliary state has been correctly prepared (preventing errors during the fault-tolerant measurement process itself from interfering with the judgment, the probability of the same error occurring twice is $\epsilon^2$). Otherwise, the quantum state is discarded and re-prepared until two consecutive measurements yield the same result.

\begin{figure}[H]
\centerline{\includegraphics[scale=0.55]{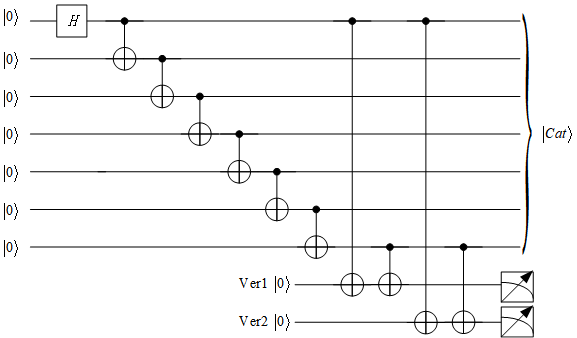}}
  \caption{Preparation and verification of the cat state}
   \label{fig:pavotcs}
\end{figure}

We first prepare a seven-qubit cat state $|Cat\rangle=\frac{1}{\sqrt{2}}(|0_C\rangle+|1_C\rangle)=\frac{1}{\sqrt{2}} (|0000000\rangle+|1111111\rangle)$, and repeat the verification twice to ensure that errors during the process do not interfere with our judgment, as shown in Figure \ref{fig:pavotcs}.

\begin{remark}
In fact, in order to reduce the number of preparations, we can directly measure three times and use the majority vote method. If two of the three results are the same, the result is considered the final judgment criterion. However, compared with our method, the quantum logical depth of this preparation method will be higher.
\end{remark}

Next, we need to use the cat state and the encoded state $|0_L\rangle$ for fault-tolerant preparation of $|\Theta \rangle$. The steps are as follows,

\begin{equation*}
    \begin{aligned}
    &|Cat\rangle|0_L\rangle =\frac{1}{\sqrt{2}}(|0_C\rangle|0_L\rangle+|1_C\rangle|0_L\rangle)\\
    &\xrightarrow{CNOT}\frac{1}{\sqrt{2}}(|0_C\rangle|0_L\rangle+|1_C\rangle|1_L\rangle)\\
    &\xrightarrow{C_{|Cat\rangle}(ZS|0_L\rangle)}\frac{1}{\sqrt{2}}(|0_C\rangle|0_L\rangle+i|1_C\rangle|1_L\rangle)\\
    &\xrightarrow{T|Cat\rangle}\frac{1}{\sqrt{2}}(|0_C\rangle|0_L\rangle+e^{i\pi/4}|1_C\rangle|1_L\rangle)\\
    &=\frac{1}{\sqrt{2}}\left(|0_C\rangle\frac{(|0_L\rangle+e^{i\pi/4}|1_L\rangle)+(|0_L\rangle-e^{i\pi/4}|1_L\rangle)}{2}+\right.\\
    &\qquad\qquad\qquad\qquad\left.|1_C\rangle\frac{(|0_L\rangle+e^{i\pi/4}|1_L\rangle)-(|0_L\rangle-e^{i\pi/4}|1_L\rangle)}{2}\right)\\
    &= \frac{1}{\sqrt{2}}\left( \frac{|0_C\rangle+|1_C\rangle}{\sqrt{2}}\frac{|0_L\rangle+e^{i\pi/4}|1_L\rangle}{\sqrt{2}}+\frac{|0_C\rangle-|1_C\rangle}{\sqrt{2}}\frac{|0_L\rangle-e^{i\pi/4}|1_L\rangle}{\sqrt{2}} \right)\\
     &\xrightarrow{H|Cat\rangle}\frac{1}{\sqrt{2}}\left( |0_C\rangle\frac{|0_L\rangle+e^{i\pi/4}|1_L\rangle}{\sqrt{2}}+ |1_C\rangle\frac{|0_L\rangle-e^{i\pi/4}|1_L\rangle}{\sqrt{2}} \right),
    \end{aligned}
\end{equation*}
\noindent where $C_{|Cat\rangle}(ZS|0_L\rangle$ means performing ZS transform on $|0_L\rangle$, controlled by $|Cat\rangle$.

\begin{figure}[H]
\centerline{\includegraphics[scale=0.7]{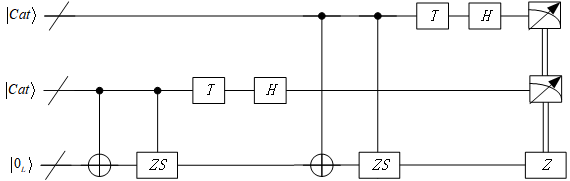}}
  \caption{Fault tolerant preparation of auxiliary state $|\Theta\rangle$}
   \label{fig:ftpoas}
\end{figure}

Based on the above steps, it is still necessary to repeat the fault-tolerant measurement twice. The preparation process of the auxiliary state $|\Theta \rangle$ is shown in Figure \ref{fig:ftpoas}.

Here, transversal implementation of the S gate requires applying ZS transform by qubit-wise. Therefore, all S gates are replaced with ZS gates for transversal implementation ($S^{\otimes 7}|1_L\rangle=-i|1_L\rangle=S^{\otimes 7}Z^{\otimes 7}|1_L\rangle$).

\begin{figure}[H]
\centerline{\includegraphics[scale=0.85]{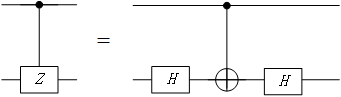}}
  \caption{Decomposition of controlled Z-gate}
   \label{fig:CZ}
\end{figure}

In Section \ref{Preliminaries}, we know that single-qubit quantum gates and CNOT gates are universal. In FTQC, we decompose controlled-Z (CZ) gates and controlled-S (CS) gates into universal gates to facilitate subsequent calculations of quantum logical depth. For CZ gates, $CZ=(I\otimes H)\cdot CNOT \cdot (I\otimes H)$, therefore, the decomposition of CZ gates is shown in Figure \ref{fig:CZ}.

\begin{figure}[H]
\centerline{\includegraphics[scale=0.8]{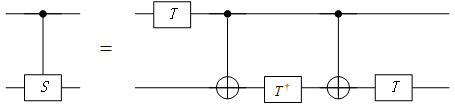}}
  \caption{Decomposition of controlled S-gate}
   \label{fig:CS}
\end{figure}

For CS gates, $CS=(T\otimes I)\cdot CNOT \cdot (I\otimes T^{\dag}) \cdot CNOT \cdot (I\otimes T)$, therefore, the decomposition of CS gates is shown in Figure \ref{fig:CS}.
\noindent Here, $T^{\dag}$ is the conjugate transpose of $T$.

\begin{figure}[H]
\centerline{\includegraphics[scale=0.75]{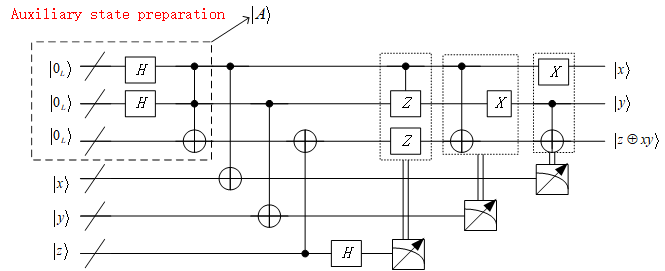}}
  \caption{Construction of fault-tolerant Toffoli gate}
   \label{fig:coftoffolig}
\end{figure}

Then, we analyze the logical depth of the fault-tolerant Toffoli gate in the encoded block and ancillary block. Its fault-tolerant implementation scheme is shown in Figure \ref{fig:coftoffolig}.

Here, $|A \rangle=\frac{1}{2}\left(|0_L0_L0_L\rangle+|0_L1_L0_L\rangle+|1_L0_L0_L\rangle+|1_L1_L1_L\rangle \right)$ is the auxiliary quantum state. Similar to the preparation of $|\Theta \rangle$, We still need the fault-tolerant measurements to prepare this auxiliary state and repeat this process twice. If the results are the same, the auxiliary state has been correctly prepared. Otherwise, the quantum state is discarded and re-prepared.

Let\begin{equation}
    \begin{aligned}
    &|B \rangle=\frac{1}{2}\left(|0_L0_L1_L\rangle+|0_L1_L1_L\rangle+|1_L0_L1_L\rangle+|1_L1_L0_L\rangle \right) \\
    &|C \rangle=\frac{1}{2}\left(|0_L0_L0_L\rangle+|0_L1_L0_L\rangle+|1_L0_L0_L\rangle+|1_L1_L0_L\rangle \right) \\
    &|D \rangle=\frac{1}{2}\left(|0_L0_L1_L\rangle+|0_L1_L1_L\rangle+|1_L0_L1_L\rangle+|1_L1_L1_L\rangle \right),
    \end{aligned}
\end{equation}
and we need to use the cat state $|Cat\rangle$ and encoded state $|0_L\rangle$ for fault-tolerant preparation of $|A\rangle$. The steps are as follows,

\begin{equation*}
    \begin{aligned}
    |Cat\rangle|0_L\rangle|0_L\rangle &|0_L\rangle
    =\frac{1}{\sqrt{2}}(|0_C\rangle|0_L0_L0_L\rangle+|1_C\rangle|0_L0_L0_L\rangle)\\
    &\xrightarrow{\overline{H}^{\otimes 3}{|0_L0_L0_L\rangle}}\frac{1}{2}(|0_C\rangle+|1_C\rangle)(|A\rangle+|B\rangle)\\
    &\xrightarrow{C_{|0_L\rangle}(Z|Cat\rangle)}\frac{1}{2}\left((|0_C\rangle+|1_C\rangle)|C\rangle+(|0_C\rangle-|1_C\rangle)|D\rangle\right)\\
    &\xrightarrow{H|Cat\rangle}\frac{1}{\sqrt{2}}\left(|0_C\rangle|C\rangle+|1_C\rangle|D\rangle\right)\\
    &\xrightarrow{Toffoli(|0_L\rangle,|0_L\rangle;|Cat\rangle)}\frac{1}{\sqrt{2}}\left(|0_C\rangle|A\rangle+|1_C\rangle|B\rangle\right),\\
    \end{aligned}
\end{equation*}
\noindent where $Toffoli(a,b;c)=c\oplus ab$, $C_{|0_L\rangle}(Z|Cat\rangle$ means performing the $Z$ transform on $|Cat\rangle$, controlled by the third $|0_L\rangle$.

\begin{figure}[H]
\centerline{\includegraphics[scale=0.75]{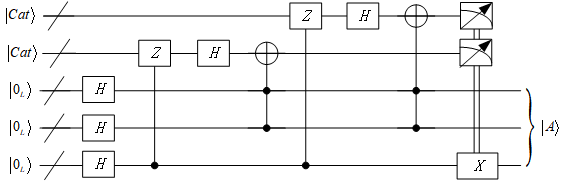}}
  \caption{Fault tolerant preparation of auxiliary state $|A\rangle$}
   \label{fig:asA}
\end{figure}

Based on the above steps, it is still necessary to repeat the fault-tolerant measurement twice. The preparation process for the ancillary state $|A\rangle$ is shown in Figure \ref{fig:asA}.

\begin{remark}
Obviously, it is also possible to perform fault-tolerant implementation based on the logical decomposition of the non-fault-tolerant structure of Toffoli, as shown in Figure \ref{fig:DOTTG}. In practice, this implementation results in a higher quantum logic depth and consumes more quantum resources.
\end{remark}

In one error-correction period, we can select the logical depth of executable fault-tolerant logic gates in an algorithm to maximize the threshold, and we provide the following definition:
\begin{definition}
A fault-tolerant quantum circuit in one error-correction period consists of an encoding circuit, a decoding circuit, and several fault-tolerant operations between them. We can select the logical depth of fault-tolerant operations between encoding and decoding, which is called a selection of one error-correction period. If this selection makes the failure probability for any quantum operation in the circuit reach the maximum threshold, it is called the optimal selection for the error-correction period (optimal error-correction period).
\end{definition}

In FTQC, we denote $p$ as the failure probability of a quantum operation, and then the equal probability of X (Y or Z) errors is $\frac{p}{3}$. In an error-correction period, if a certain physical qubit only occurs X errors and the logical depth (excluding perfect quantum operations) is $r_x$, then its error probability is $1-(1-\frac{p}{ 3})^{r_x} \approx \frac{p}{3}{r_x}$. Similarly, the probability of Z or Y errors is $\frac{p}{3}{r_z}$ or $\frac{p}{3}{r_y}$. Therefore, the failure probability of this physical qubit is $\frac{p}{3}(r_x+r_y+r_z)$. Let $R=\lceil \frac{r_x+r_y+r_z}{3} \rceil$, then its failure probability is $Rp$. If there are two or more same type of errors in an encoded block, an unrecoverable error will occur, and the failure probability is $\mathcal{O}(p^2) \approx cp^2$, where $c$ is the number of all possible failure point pairs. When concatenating one level based on Steane code, the probability of an unrecoverable error on a logical qubit is $cp^2$. When concatenating $k$ levels, the probability of an unrecoverable error on a logical qubit is $\frac{1}{c}(cp)^{2^k}$, where $c$ is related to $k$.

In a quantum circuit without fault-tolerant structure, we denote $r$ as the logical depth of a certain logical qubit before the measurement during the algorithm execution, and $x$ as the logical depth of a physical qubit of the executable algorithm in an error-correction period. $\Delta=\frac{r_0}{x}$ represents the number of error-correction periods, where $r_0$ is the logical depth of the physical qubit before the fault-tolerant measurement during the algorithm execution. After performing a complete algorithm, the error probability on a logical qubit is $rp$. Therefore, we have a threshold condition
\begin{equation}\label{eq7}
    \frac{1}{c}(cp)^{2^k} \cdot \Delta  \leq rp   ,
\end{equation}

This means that by using a $k$-level Steane code, the error probability of one qubit can be reduced from $rp$ to $\frac{1}{c}(cp)^{2^k} \cdot \Delta$. This threshold condition in formula \ref{eq7} is equivalent to
\begin{equation}
 p \leq \frac{(r/\Delta)^{\frac{1}{2^k-1}}}{c}  ,
\end{equation}
So we define the threshold $p_{th}$ as follows:
\begin{equation}
\label{eq9}
    p_{th}=\frac{(r/\Delta)^{\frac{1}{2^k-1}}}{c}  ,
\end{equation}
where $c$ is not only related to $k$, but also related to $x$.

From formula \ref{eq9}, we can know that the threshold $p_{th}$ depends on the number $k$ of concatenated levels, the logical depth $r$ of the logical qubit, and the number $\Delta$ of error-correction periods. In fact, $p_{th}$ depends only on $k$, $r$, and $x$. If we fix the value of $k$, then the value of $p_{th}$ depends only on the values of $r$ and $x$, and we can obtain the optimal error-correction period.

Let the average logical depth (excluding perfect quantum operations) of the seven physical qubits during the encoding and decoding process when concatenating one level Steane code be $R_1$, $R_2$, $R_3$, $R_4$, $R_5$, $R_6$, $R_7$, respectively. Since we use $|Steane\rangle$ for the syndrome measurement and recovery process, the logical depth of the seven physical qubits in this process is all $\gamma$.

In our scheme, $R_2=R_3$, $R_6=R_7$. Since $c$ is related to $k$ and $x$, we propose an algorithm to compute $c$ and $p_{th}$. The algorithm is as follows: 
\begin{algorithm}[H]
		\caption{\textbf{Algorithm for solving coefficient $c$ and $p_{th}$}}
		\label{Alg:1}
		\begin{algorithmic}[1]
            \State   $\text{list}_0$=[$R_1,R_2,R_3,R_4,R_5,R_6,R_7$];
            \State  \textbf{for}  $k \leftarrow 1$ $\textbf{to}$ $\infty$ \textbf{do}
            \State \ \ \ \ $\text{list}_1$=[ ];
            \State \ \ \ \  \textbf{if} $k==1$ \textbf{then}
            \State \ \ \ \ \ \ \ \  $\text{list}_0$=$\text{list}_0$;
            \State \ \ \ \ \textbf{else}
            \State \ \ \ \ \ \ \ \  \textbf{for}  $i \leftarrow 0$ $\textbf{to}$ len($\text{list}_0$) \textbf{do}
            \State \ \ \ \ \ \ \ \ \ \ \ \ $\text{list}_1$.append($\text{list}_0$[i]+$R_1$);
            \State \ \ \ \ \ \ \ \ \ \ \ \ \textbf{for}  $j \leftarrow 1$ $\textbf{to}$ 7 \textbf{do}
             \State \ \ \ \ \ \ \ \ \ \ \ \ \ \ \ \ $\text{list}_1$.append($\text{list}_0$[j]);
            \State \ \ \ \ \ \ \ \ $\text{list}_0$=$\text{list}_1$;
            \State \ \ \ \  \textbf{for}  $r \leftarrow 1$ $\textbf{to}$ $\infty$ \textbf{do}
            \State \ \ \ \ \ \ \ \ $\text{list}_{p_{th}}$=[ ];
            \State \ \ \ \ \ \ \ \ \textbf{for}  $x \leftarrow 1$ $\textbf{to}$ $\infty$ \textbf{do}
            \State \ \ \ \ \ \ \ \ \ \ \ \ $c_0=0$;
            \State \ \ \ \ \ \ \ \ \ \ \ \  \textbf{for}  $s$ in range (0,$7^k$,7) \textbf{do}
            \State \ \ \ \ \ \ \ \ \ \ \ \ \ \ \ \  $c_0 = 2(\text{list}_0[s]+\gamma x)(R_2+\gamma x)+(\text{list}_0[s]+\gamma x)(R_4+\gamma x)+(\text{list}_0[s]+\gamma x)(R_5+\gamma x)+2(\text{list}_0[s]+\gamma x)(R_6+\gamma x)+(R_2+\gamma x)^2+2(R_2+\gamma x)(R_4+\gamma x)+2(R_2+\gamma x)(R_5+\gamma x)+(R_4+\gamma x)(R_5+\gamma x)+4(R_2+\gamma x)(R_6+\gamma x)+2(R_4+\gamma x)(R_6+\gamma x)+2(R_5+\gamma x)(R_6+\gamma x)+(R_6+\gamma x)^2+c_0$;
            \State \ \ \ \ \ \ \ \ \ \ \ \  $c = c_0/(7^{k-1})$;
            \State \ \ \ \ \ \ \ \ \ \ \ \  $p_{th} = (rx/r_0)^{\frac{1}{2^k-1}} / c$;
            \State \ \ \ \ \ \ \ \ \ \ \ \     $\text{list}_{p_{th}}.append(p_{th})$;
            \State \ \ \ \ \ \ \ \ max\_value = max($\text{list}_{p_{th}}$);
            \State \ \ \ \ \ \ \ \ max\_index = $\text{list}_{p_{th}}$.index(max\_value) + 1 ;
            \State \ \ \ \ \ \ \ \ return (k,r,max\_index,max\_value)
		\end{algorithmic}
\end{algorithm}

\subsection{Simulation and Analysis}
For the quantum gate operations that can be directly implemented transversally, the quantum operation is performed on the original data block, so $r_0=r$, and the formula \ref{eq9} becomes
\begin{equation}
    p_{th}=  \frac{x^{\frac{1}{2^k-1}}}{c}  .
\end{equation}

We simulate the fault-tolerant Steane code for the data block in Section \ref{fault-tolerant implementation based on Steane code}, $R_1=\lceil \frac{9+5+5}{3}\rceil=7$, and similarly, $R_2=R_3=13$, $R_4= 15$, $R_5=14$, $R_6=R_7=10$; for the auxiliary block $|0_L\rangle$, $R_1=\lceil \frac{8+0+8}{3}\rceil=6$, and similarly, $R_2=R_3=R_4=8$, $R_5=7$, $R_6=R_7=6$. The logical depth of the seven physical qubits in the syndrome measurement and recovery process is $\gamma=4$ (assuming that the single-qubit quantum gate operation used for error correction is perfect).

For the original data block and auxiliary block, to gain a clearer understanding of the relationship among concatenated levels $k$, the logical depth $x$ of a physical qubit in the executable algorithm during an error-correction period, and the maximum threshold $max(p_{th})$, we list these in Table \ref{tab:1}.
\begin{table}[H]
\caption{The relationship among $k,x,p_{th}$ for data block and auxiliary block}
\label{tab:1}
\begin{subtable}{.5\linewidth}
\caption{$k,x,p_{th}$ in the data block}
\centering
\begin{adjustbox}{max width=\textwidth}
\begin{tabular}{|>{\centering\arraybackslash}p{0.5cm}|>{\centering\arraybackslash}p{0.5cm}|c|}
\hline
\textbf{k} & \textbf{x} & \bm{$max(p_{th})$} \\ \hline
1 & 3 & 2.545392838961480e-04 \\ \hline
2 & 1 & 1.581849407936365e-04 \\ \hline
3 & 1 & 1.541452488659314e-04 \\ \hline
4 & 1 & 1.535849320196374e-04 \\ \hline
5 & 1 & 1.535052191135160e-04 \\ \hline
6 & 1 & 1.534938383096437e-04 \\ \hline
7 & 1 & 1.534922126182756e-04 \\ \hline
8 & 1 & 1.534919803794627e-04 \\ \hline
9 & 1 & 1.534919472025467e-04 \\ \hline
10 & 1 & 1.534919424629885e-04 \\ \hline
$\cdots$ & $\cdots$ & $\cdots$ \\ \hline
\end{tabular}
\end{adjustbox}
\end{subtable}
\begin{subtable}{.5\linewidth}
\caption{$k,x,p_{th}$ in the auxiliary block}
\centering
\begin{adjustbox}{max width=\textwidth}
\begin{tabular}{|>{\centering\arraybackslash}p{0.5cm}|>{\centering\arraybackslash}p{0.5cm}|c|}
\hline
\textbf{k} & \textbf{x} & \bm{$max(p_{th})$} \\ \hline
1 & 2 & 4.235493434985176e-04 \\ \hline
2 & 1 & 3.325573661456601e-04 \\ \hline
3 & 1 & 3.253090435914119e-04 \\ \hline
4 & 1 & 3.242992819087329e-04 \\ \hline
5 & 1 & 3.241555417366799e-04 \\ \hline
6 & 1 & 3.241350178274260e-04 \\ \hline
7 & 1 & 3.241320860525464e-04 \\ \hline
8 & 1 & 3.241316672318930e-04 \\ \hline
9 & 1 & 3.241316074004594e-04 \\ \hline
10 & 1 & 3.241315988531136e-04 \\ \hline
$\cdots$ & $\cdots$ & $\cdots$ \\ \hline
\end{tabular}
\end{adjustbox}
\end{subtable}
\end{table}

From this, we can select the optimal error-correction period under different concatenated levels and obtain the maximum threshold.

However, in fact, the error-correction codes in quantum computers cannot be infinitely concatenated. Knill proposed a simple model for FTQC and introduced the currently highest threshold of $10^{-2}$ \cite{knill2005quantum}. On this basis, \cite{yang2007upper} and \cite{yang2008decoherent} proposed an upper bound on the number of gates for a single qubit in an error-correction period based on the decoherence limit of coherent field-driven FTQC.

In \cite{yang2013full}, combined with the FTQC threshold Theorem \ref{theorem1}, a theoretical framework for the permitted logical depth in quantum computation is proposed. \cite{yang2020effect} further developed the results from \cite{yang2013full}, applying the full-quantum theory of atom-field interaction to the ion trap quantum computation schemes. This work provides the first full-quantum operation results for CNOT gates and subsequently provides the error probability after multiple CNOT gate operations. Ultimately, it proposed the concept of the permitted logical depth for the ion trap FTQC. Based on these theories, we introduce the definition of the permitted logical depth, as follows: 
\begin{definition}
In an error-correction period of FTQC, the maximum number of physical quantum operations permitted on a physical qubit is called the permitted logical depth of FTQC. Its value is approximately of the order $10^2$. 
\end{definition}

\begin{figure}[H]
    \centering 
    \includegraphics[scale=0.6]{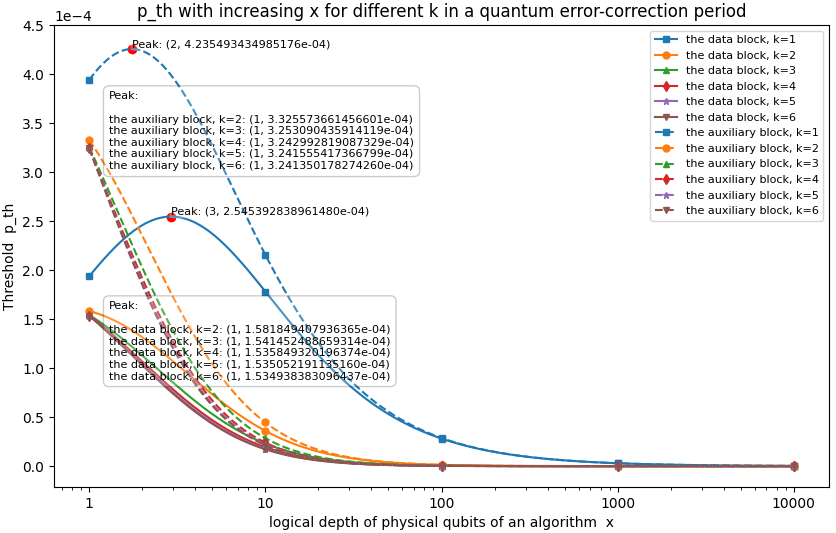}
    \caption{The relationship among $k,x,p_{th}$ for data block and auxiliary block}
    \label{fig:trfdq}
\end{figure}

This is independent of future technological advancements and is an inherent physical limitation imposed by fundamental physical laws that cannot be overcome. Therefore, the number of concatenated levels cannot increase indefinitely. Combined with Table \ref{tab:1} and the above theory, our scheme can approximately concatenate 6 levels. Using Algorithm \ref{Alg:1}, we plot the relationship curve between $p_{th}$ and $x$ within an error-correction period, as shown in Figure \ref{fig:trfdq}.

It can be seen from Figure \ref{fig:trfdq} and Table \ref{tab:1} that only when $k=1$, i.e., without using concatenated codes, the logical depth of the physical qubits executing the algorithm within the error-correction period is 3 in a data block (is 2 in an auxiliary block), the threshold can reach the maximum; for other concatenated levels, the threshold reaches the maximum when the logical depth of the physical qubits executing the algorithm within one error-correction period is 1. Since the number of operations in the encoding and decoding process of the auxiliary block is less than that of the data block, its threshold is significantly higher than that of the data block according to Table \ref{tab:1} and Figure \ref{fig:trfdq}. The maximum threshold of the entire FTQC process should be the maximum threshold in the data block. No matter how many levels are concatenated, as $x$ increases, the maximum threshold becomes smaller and smaller; as $k$ increases, the maximum threshold gradually decreases but the order of magnitude is $\mathcal{O}(10^{-4} )$. This result is consistent with the threshold Theorem \ref{theorem1} and the widely recognized threshold \cite{nielsen2010quantum}.

Compared with the technique of introducing malignant set counting in \cite{aliferis2005quantum} to analyze the malignant pair combinations of failure places in quantum logic gates, our method considers all the failure point pairs during the encoding and decoding process; compared with \cite{quan2022implementation} which analyzes error propagation based on stabilizers and flag qubits, our method uses the Steane state for fault-tolerant syndrome measurements, reducing the number of CNOT gates, and improves the scheme for detecting X-type errors using flag qubits, preventing the propagation of errors caused by Z-type errors on the target qubits of introduced auxiliary CNOT gates. Additionally, we analyze all quantum gates in the encoding and decoding processes, including the introduced auxiliary gates, that may cause errors.

According to Remark \ref{rm4}, if the original data qubits pass through the T gate or Toffoli gate, then the fault-tolerant measurement is required, causing collapse. At this time, the logical operations will be transferred to the auxiliary block and the operation results will be stored in the auxiliary block. In this case, $r_0=r-1+r',\Delta=\frac{r-1+r'}{x}$, and the formula \ref{eq9} becomes
\begin{equation}
    p_{th}=\frac{  (\frac{rx}{r-1+r'})^{\frac{1}{2^k-1}}}{c},  
\end{equation}
where $r'$ represents the logical depth of the fault-tolerant T gate or the fault-tolerant Toffoli gate in the auxiliary block. For fault-tolerant T gates, $r'=20$; for fault-tolerant Toffoli gates, $r_1'=19, r_2'=17, r_3'=8$ respectively represent the logic depth of the first two control qubits and the target qubit.

\begin{figure}[H]
    \centering 
    \includegraphics[scale=0.6]{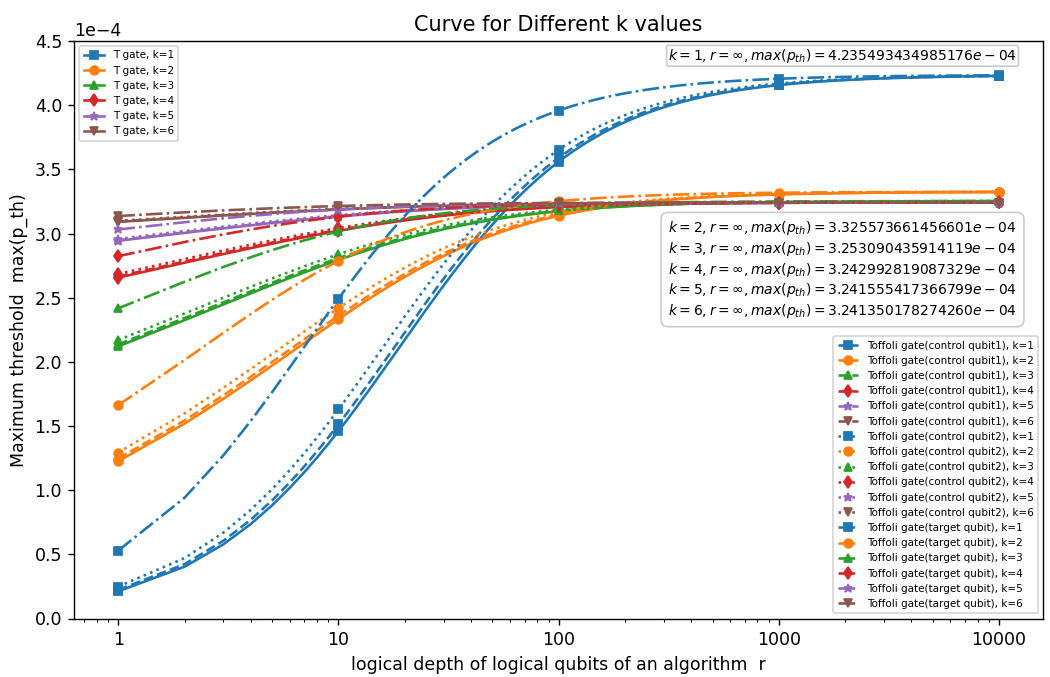}
    \caption{The relationship among $k,x,max(p_{th})$ for auxiliary block}
    \label{fig:trfaq}
\end{figure}

In fact, for different $r$, the logical depth $x$ of the physical qubits executing the algorithm in an error-correction period with the maximum threshold under the same concatenated level is fixed (as known from step 17 in Algorithm \ref{Alg:1}). Therefore, the maximum threshold is only related to $r$ and $k$. We plot the curve between their maximum threshold $max(p_{th})$ and $r$ under different concatenated levels $k$, as shown in Figure \ref{fig:trfaq}.

Since $k, r$ can take any positive integer, we list some representative data to show the relationship among $k, r, max(p_{th})$, see Table \ref{tab:2}.
\begin{table*}[h!]
\caption{The relationship among $k,x,max(p_{th})$ for auxiliary block}
\label{tab:2}
\begin{subtable}{1\linewidth}
\caption{$k,x,max(p_{th})$ for T gate}
\centering
\begin{adjustbox}{max width=0.85\textwidth}
\begin{tabular}{|c|c|c|c|c|c|c|c|c|}
\hline
\multicolumn{1}{|c|}{\textbf{k}} & \textbf{r} & \bm{$max(p_{th})$} & \multicolumn{1}{c|}{\textbf{k}} & \textbf{r} & \bm{$max(p_{th})$} & \multicolumn{1}{c|}{\textbf{k}} & \textbf{r} & \bm{$max(p_{th})$} \\ \hline
\multirow{6}{*}{1} & 1 & 2.117746717492588e-05 & \multirow{6}{*}{2} & 1 & 1.225151811985496e-04 & \multirow{6}{*}{3} & 1 & 2.120482579274038e-04 \\ \cline{2-3} \cline{5-6} \cline{8-9} 
 & 10 & 1.460514977581095e-04 &  & 10 & 2.332028780560102e-04 &  & 10 & \multicolumn{1}{c|}{2.794082852232359e-04} \\ \cline{2-3} \cline{5-6} \cline{8-9} 
 & 100 & 3.559238180659812e-04 &  & 100 & 3.138226254720990e-04 &  & 100 & \multicolumn{1}{c|}{3.173245799080812e-04} \\ \cline{2-3} \cline{5-6} \cline{8-9} 
 & 1000 & 4.156519563282803e-04 &  & 1000 & 3.304774598767125e-04 &  & 1000 & \multicolumn{1}{c|}{3.244355203675500e-04} \\ \cline{2-3} \cline{5-6} \cline{8-9} 
 & 10000 & 4.227461258593848e-04 &  & 10000 & 3.323470128717182e-04 &  & 10000 & \multicolumn{1}{c|}{3.252208411591097e-04} \\ \cline{2-3} \cline{5-6} \cline{8-9} 
 & $\infty$ & 4.235493434985176e-04 &  & $\infty$ & 3.325573661456601e-04 &  & $\infty$ & \multicolumn{1}{c|}{3.253090435914119e-04} \\ \hline
\multicolumn{1}{|c|}{\textbf{k}} & \textbf{r} & \bm{$max(p_{th})$} & \multicolumn{1}{c|}{\textbf{k}} & \textbf{r} & \bm{$max(p_{th})$} & \multicolumn{1}{c|}{\textbf{k}} & \textbf{r} & \bm{$max(p_{th})$} \\ \hline
\multirow{6}{*}{4} & 1 & 2.655893487303872e-04 & \multirow{6}{*}{5} & 1 & 2.942962587328901e-04 & \multirow{6}{*}{6} & 1 & 3.090826895278016e-04 \\ \cline{2-3} \cline{5-6} \cline{8-9}
 & 10 & \multicolumn{1}{c|}{3.020782480236627e-04} &  & 10 & 3.132112741262893e-04 &  & 10 & 3.187031101529598e-04 \\ \cline{2-3} \cline{5-6} \cline{8-9} 
 & 100 & \multicolumn{1}{c|}{3.205601428289243e-04} &  & 100 & 3.223416702370663e-04 &  & 100 & 3.232412624832499e-04 \\ \cline{2-3} \cline{5-6} \cline{8-9} 
 & 1000 & \multicolumn{1}{c|}{3.238926116780298e-04} &  & 1000 & 3.239587893449915e-04 &  & 1000 & 3.240381943606451e-04 \\ \cline{2-3} \cline{5-6} \cline{8-9} 
 & 10000 & \multicolumn{1}{c|}{3.242582455709016e-04} &  & 10000 & 3.241356935969836e-04 &  & 10000 & 3.241252517489947e-04  \\ \cline{2-3} \cline{5-6} \cline{8-9} 
 & $\infty$ & \multicolumn{1}{c|}{3.242992819087329e-04} &  & $\infty$ & 3.241555417366799e-04 &  & $\infty$ & \multicolumn{1}{c|}{3.241350178274260e-04} \\ \hline
\end{tabular}
\end{adjustbox}
\end{subtable}
\begin{subtable}{1\linewidth}
\caption{$k,x,max(p_{th})$ for Toffoli gate (target qubit)}
\centering
\begin{adjustbox}{max width=0.85\textwidth}
\begin{tabular}{|c|c|c|c|c|c|c|c|c|}
\hline
\multicolumn{1}{|c|}{\textbf{k}} & \textbf{r} & \bm{$max(p_{th})$} & \multicolumn{1}{c|}{\textbf{k}} & \textbf{r} & \bm{$max(p_{th})$} & \multicolumn{1}{c|}{\textbf{k}} & \textbf{r} & \bm{$max(p_{th})$} \\ \hline
\multirow{6}{*}{1} & 1 & 5.294366793731470e-05 & \multirow{6}{*}{2} & 1 & 1.662786830728301e-04 & \multirow{6}{*}{3} & 1 & 2.417036904907203e-04 \\ \cline{2-3} \cline{5-6} \cline{8-9}
 & 10 & 2.491466726461868e-04 &  & 10 & 2.786443649940462e-04 &  & 10 & \multicolumn{1}{c|}{3.015607830486628e-04} \\ \cline{2-3} \cline{5-6} \cline{8-9} 
 & 100 & 3.958405079425398e-04 &  & 100 & 3.251411813479597e-04 &  & 100 & \multicolumn{1}{c|}{3.221799088505769e-04} \\ \cline{2-3} \cline{5-6} \cline{8-9} 
 & 1000 & 4.206051077443075e-04 &  & 1000 & 3.317850005371942e-04 &  & 1000 & \multicolumn{1}{c|}{3.249850293133916e-04} \\ \cline{2-3} \cline{5-6} \cline{8-9} 
 & 10000 & 4.232530663520711e-04 &  & 10000 & 3.324798056189911e-04 &  & 10000 & \multicolumn{1}{c|}{3.252765256929118e-04} \\ \cline{2-3} \cline{5-6} \cline{8-9}
 & $\infty$ & 4.235493434985176e-04 &  & $\infty$ & 3.325573661456601e-04 &  & $\infty$ & \multicolumn{1}{c|}{3.253090435914119e-04} \\ \hline
\multicolumn{1}{|c|}{\textbf{k}} & \textbf{r} & \bm{$max(p_{th})$} & \multicolumn{1}{c|}{\textbf{k}} & \textbf{r} & \bm{$max(p_{th})$} & \multicolumn{1}{c|}{\textbf{k}} & \textbf{r} & \bm{$max(p_{th})$} \\ \hline
\multirow{6}{*}{4} & 1 & 2.823189225421760e-04 & \multirow{6}{*}{5} & 1 & 3.031248322460440e-04 & \multirow{6}{*}{6} & 1 & 3.136109302804428e-04 \\ \cline{2-3} \cline{5-6} \cline{8-9}
 & 10 & \multicolumn{1}{c|}{3.130276678412809e-04} &  & 10 & 3.186541761167236e-04 &  & 10 & 3.214164004069907e-04 \\ \cline{2-3} \cline{5-6} \cline{8-9} 
 & 100 & \multicolumn{1}{c|}{3.228397991964612e-04} &  & 100 & 3.234488317485713e-04 &  & 100 & 3.237871009174483e-04 \\ \cline{2-3} \cline{5-6} \cline{8-9} 
 & 1000 & \multicolumn{1}{c|}{3.241485045352345e-04} &  & 1000 & 3.240826085281899e-04 &  & 1000 & 3.240991302796217e-04 \\ \cline{2-3} \cline{5-6} \cline{8-9} 
 & 10000 & \multicolumn{1}{c|}{3.242841535895349e-04} &  & 10000 & 3.241482247386771e-04 &  & 10000 & 3.241314176071593e-04 \\ \cline{2-3} \cline{5-6} \cline{8-9} 
 & $\infty$ & \multicolumn{1}{c|}{3.242992819087329e-04} &  & $\infty$ & 3.241555417366799e-04 &  & $\infty$ & \multicolumn{1}{c|}{3.241350178274260e-04} \\ \hline
\end{tabular}
\end{adjustbox}
\end{subtable}
\end{table*}

It can be seen from Figure \ref{fig:trfaq} and Table \ref{tab:2} that while ensuring that the maximum threshold by selecting the logical depth $x$ of the physical qubits executing the algorithm within the error-correction period, the maximum threshold increases with the increase of concatenated levels $k$ when $r$ is the same; the maximum threshold $max(p_{th})$ is also related to $r$, and increases with the increase of $r$. However, there is a limit value, namely, the maximum threshold $max(p_{th})$ corresponding to different concatenated levels $k$ in Table \ref{tab:1} (b).

Our simulation results show that the selection of the optimal error-correction period depends on the values of $k$ and $x$. In order to increase the threshold, one can increase the number of concatenated levels while simultaneously increasing the value of $r$. For example, increasing the number of repetitions in preparing the ancillary state can increase $r$.

\subsection{Quantum security}
The operating principles of quantum computers are more complex than those of classical computers. During the operation, quantum computers can introduce many unnecessary errors and interferences, which are constrained by the fundamental physical laws and are not eliminated with technological advancements. To ensure the reliability and accuracy of quantum computers, running actual quantum algorithms is based on the FTQC schemes. The running time of quantum algorithms depends on the running time of specific physical components.

The CNOT gates can be regarded as the key to realizing quantum computation. The computational complexity of an algorithm is reduced to the number of CNOT gates \cite{knill2005quantum}. Considering the universal quantum gates in an ion trap quantum computer, the running time of a single-qubit quantum gate is much less than that of a double-qubit quantum gate \cite{yang2013full,yang2020effect}, so it is sufficient to analyze the CNOT gates. In our FTQC scheme, as shown in Figure \ref{fig:ftscbfq}, an encoded block requires at least 36 CNOT gates, and we introduce the "flag qubits" scheme, which requires 16 CNOT gates. Combined with our simulation results, for quantum gates that can be directly implemented transversally, an encoded block without using concatenated codes in one error-correction period requires at least 52 CNOT gates. In Figure \ref{fig:cofttg}, it requires at least 7 CNOT gates; in Figure \ref{fig:pavotcs}, it requires at least 10 CNOT gates; and in Figure \ref{fig:ftpoas}, combined with Figure \ref{fig:CZ} and Figure \ref{fig:CS}, it requires at least 56 CNOT gates. Therefore, for the T gate, as shown in Figure \ref{fig:cofttg}, \ref{fig:pavotcs} and \ref{fig:ftpoas}, it requires at least 135 CNOT gates (including one encoded block $|\Theta\rangle$, 52 CNOT gates; two cat states, 20 CNOT gates), and an additional 40 CNOT gates for the $|0_L\rangle$ auxiliary block with the Flag scheme in Figure \ref{fig:Flag-X}, so at least 175 CNOT gates are required in one error-correction period without using concatenated codes. In Figure \ref{fig:coftoffolig}, it requires at least 42 CNOT gates; in Figure \ref{fig:DOTTG}, a Toffoli gate can be decomposed into 6 CNOT gates, and in Figure \ref{fig:asA}, combined with Figure \ref{fig:CZ} and Figure \ref{fig:CS}, it requires at least 98 CNOT gates. Therefore, for the Toffoli gate, as shown in Figures \ref{fig:DOTTG}, \ref{fig:pavotcs}, \ref{fig:coftoffolig}, and \ref{fig:asA}, it requires at least 316 CNOT gates (including three encoded blocks $|x\rangle$,$|y\rangle$,$|z\rangle$, 156 CNOT gates; two cat states, 20 CNOT gates) and involves three $|0_L\rangle$ auxiliary blocks, so at least 436 CNOT gates are required in one error-correction period without using concatenated codes.

The above analysis is only for the case where the logical 1-qubit utilizes the least number of CNOTs in one error-correction period without using concatenated codes, which has reached $\mathcal{O}(10^2)$. The running time of a quantum computer is closely related to the size of a quantum algorithm, including its length of input qubits and the number of logic gates. When ensuring the maximum threshold, as the number of concatenated levels increases, the number of CNOT gates increases exponentially. Yang et al. \cite{yang2013post} proposed the lower bound of the time required for a single CNOT operation in an ion trap quantum computer for the first time by analyzing the phonon speed, which is $2.85 \times 10^{-4}$s. Therefore, in actual quantum computers, many known quantum algorithms require a long time to run, even exceeding a meaningful time frame.

We study the computational theoretical limits of quantum computers that can provide design guidelines in quantum computation environments, which lays the foundation for the idea of active defense. For the ion trap quantum computer that is most likely to be realized first, generally, the number of operations on the same physical qubit cannot exceed its permitted logical depth of the order $10^2$ \cite{yang2020effect}, and the running time may make the quantum attack algorithm unable to be successfully executed. The system using this idea will not lose security with the development of future quantum algorithms and can achieve the effect of real active defense and information security.

\section{Discussion}\label{discussion}
In this paper, we consider the results caused by different types of errors on each quantum gate operation. However, there are still two insurmountable difficulties. The first difficulty is that regardless of whether auxiliary quantum gates are introduced, the last quantum operation cannot be guaranteed that errors will not occur, and we cannot detect them. The second difficulty is errors that cannot be detected due to the structure of the encoding itself, such as Z errors on information qubits, which can be considered as a change in information and thus cannot be detected. Except for these two cases, our scheme ensures that all possible errors before the last quantum operation of the decoding process can be detected. For single-qubit quantum gates used for correction operation during the last round of recovery, their error probability may be reduced through technological improvements, but there is no guarantee that errors will not occur. FTQC will certainly not be perfect, but our scheme can make it as perfect as possible.

Our algorithm is a more accurate estimation, analyzing the respective logical depth of the seven physical qubits in an encoded block after multiple-level concatenation. This is because, as the number of concatenated levels increases, the logical depth of the data qubits differs significantly from that of other redundant qubits. If we directly use the maximum logical depth for computation, it may lead to significant errors. However, our algorithm is still an estimation method. We provide a threshold analysis framework, and the threshold can still be analyzed more accurately in the future. The error probability in single-qubit gates is generally lower than that in CNOT gates, and different FTQC schemes will have different thresholds.

Given an FTQC scheme, the logical depth of the encoding and decoding, along with the optimal error-correction period to reach the maximum threshold, are fixed. Therefore, the only way to increase the threshold is by increasing the logical depth of physical qubits in the auxiliary block before fault-tolerant measurements. For the T gate or Toffoli gate, there may be better fault-tolerant schemes in the future. For a specific fault-tolerant preparation method, the threshold can be increased by increasing the number of verifications when preparing auxiliary states. However, this will consume more unnecessary quantum resources, so the trade-off between maximum threshold and quantum resources will also be a worthwhile research problem.

\section{Conclusion}\label{conclusion}
We propose a fault-tolerant encoding and decoding scheme based on Steane code. It uses Steane states for fault-tolerant syndrome measurements, combined with the results of measuring redundant qubits during the decoding process to realize the fault-tolerant Steane code. However, due to error propagation, the same measurement results may correspond to different errors. To reduce the interference, we introduce the "flag qubits" scheme and provide its usage conditions.

We consider all possible errors and error propagation caused by quantum gates in one error-correction period, detecting and correcting these errors. Theoretical analysis shows that any distinct error will result in a unique syndrome, allowing for accurate differentiation and correction. In addition, we also provide the fault-tolerant implementation scheme for the universal quantum gate set, including fault-tolerant preparation and verification of auxiliary states. For the first time, we consider the fault tolerance for all processes of FTQC. We propose a more accurate threshold estimation algorithm, according to which the optimal error-correction period can be obtained, as well as a method to increase the threshold. We obtain the following conclusions:

\begin{enumerate}
    \item Regardless of the number of concatenated levels, as the number of algorithmic fault-tolerant quantum gates increases, the maximum threshold decreases. As the number of concatenated levels increases, the optimal selection for the error-correction period is to perform only one fault-tolerant quantum gate of the algorithm when using concatenated codes, resulting in the maximum threshold. The maximum threshold gradually decreases with the increase in concatenated levels, but it remains the order $\mathcal{O}(10^{-4})$, consistent with the widely recognized thresholds \cite{nielsen2010quantum}.
    
    \item Under the same number of concatenated levels, the threshold can be increased by increasing the logical depth of physical qubits in the auxiliary block before fault-tolerant measurements. For example, increasing the number of repeated measurements when preparing auxiliary states. Under the same logical depth of logical qubits in the algorithm, the threshold can be increased by increasing the number of concatenated levels. However, the limit value of increasing the threshold is the maximum threshold within the optimal error-correction period after fixing the number of concatenated levels in the auxiliary block.
\end{enumerate}

The simulation results show the effectiveness of this method, which is consistent with the threshold Theorem \ref{theorem1}. This fault-tolerant encoding and decoding scheme can be extended to other CSS codes to improve the reliability of FTQC. In future work, we can also fully consider the characteristics of surface code stabilizers and extend this method to the most potential error-correction code in quantum computation—surface code. In addition, we study the computational theoretical limits of quantum computers and analyze the actual running time of quantum algorithms from the perspectives of attack and active defense based on our FTQC scheme, thereby assessing the security of a system.

\section*{Acknowledgements}
This work was supported by Beijing Natural Science Foundation (Grant No.4234084).

 \bibliographystyle{elsarticle-num} 
 \bibliography{cas-refs}

\begin{thebibliography}{10}
\expandafter\ifx\csname url\endcsname\relax
  \def\url#1{\texttt{#1}}\fi
\expandafter\ifx\csname urlprefix\endcsname\relax\def\urlprefix{URL }\fi
\expandafter\ifx\csname href\endcsname\relax
  \def\href#1#2{#2} \def\path#1{#1}\fi

\bibitem{shor1994algorithms}
P.~W. Shor, Algorithms for quantum computation: discrete logarithms and
  factoring, in: Proceedings 35th annual symposium on foundations of computer
  science, Ieee, 1994, pp. 124--134.

\bibitem{shor1994polynomial}
P.~W. Shor, Polynomial-time algorithms for prime factorization and discrete
  logarithms, in: Proceedings of the 35th Annual Symposium on Foundations of
  Computer Science, Vol. 124, 1994, pp. 124--134.

\bibitem{grover1996fast}
L.~K. Grover, A fast quantum mechanical algorithm for database search, in:
  Proceedings of the twenty-eighth annual ACM symposium on Theory of computing,
  1996, pp. 212--219.

\bibitem{grover1997quantum}
L.~K. Grover, Quantum mechanics helps in searching for a needle in a haystack,
  Physical Review Letters 79~(2) (1997) 325--328.

\bibitem{landauer1995quantum}
R.~Landauer, Is quantum mechanics useful?, Philosophical Transactions of the
  Royal Society of London. Series A: Physical and Engineering Sciences
  353~(1703) (1995) 367--376.

\bibitem{chuang1995quantum}
I.~L. Chuang, R.~Laflamme, P.~W. Shor, W.~H. Zurek, Quantum computers,
  factoring, and decoherence, Science 270~(5242) (1995) 1633--1635.

\bibitem{unruh1995maintaining}
W.~G. Unruh, Maintaining coherence in quantum computers, Physical Review A
  51~(2) (1995) 992.

\bibitem{landauer1996physical}
R.~Landauer, The physical nature of information, Physics letters A 217~(4-5)
  (1996) 188--193.

\bibitem{zurek1991decoherence}
W.~H. Zurek, Decoherence and the transition from quantum to classical, Physics
  today 44~(10) (1991) 36--44.

\bibitem{bernstein1993quantum}
E.~Bernstein, U.~Vazirani, Quantum complexity theory, in: Proceedings of the
  twenty-fifth annual ACM symposium on Theory of computing, 1993, pp. 11--20.

\bibitem{bennett1997strengths}
C.~H. Bennett, E.~Bernstein, G.~Brassard, U.~Vazirani, Strengths and weaknesses
  of quantum computing, SIAM journal on Computing 26~(5) (1997) 1510--1523.

\bibitem{gottesman2002introduction}
D.~Gottesman, An introduction to quantum error correction, in: Proceedings of
  Symposia in Applied Mathematics, Vol.~58, 2002, pp. 221--236.

\bibitem{shor1995scheme}
P.~W. Shor, Scheme for reducing decoherence in quantum computer memory,
  Physical review A 52~(4) (1995) R2493.

\bibitem{steane1996error}
A.~M. Steane, Error correcting codes in quantum theory, Physical Review Letters
  77~(5) (1996) 793.

\bibitem{steane1996multiple}
A.~Steane, Multiple-particle interference and quantum error correction,
  Proceedings of the Royal Society of London. Series A: Mathematical, Physical
  and Engineering Sciences 452~(1954) (1996) 2551--2577.

\bibitem{calderbank1996good}
A.~R. Calderbank, P.~W. Shor, Good quantum error-correcting codes exist,
  Physical Review A 54~(2) (1996) 1098.

\bibitem{shor1996fault}
P.~W. Shor, Fault-tolerant quantum computation, in: Proceedings of 37th
  conference on foundations of computer science, IEEE, 1996, pp. 56--65.

\bibitem{preskill1998fault}
J.~Preskill, Fault-tolerant quantum computation, in: Introduction to quantum
  computation and information, World Scientific, 1998, pp. 213--269.

\bibitem{gottesman1998theory}
D.~Gottesman, Theory of fault-tolerant quantum computation, Physical Review A
  57~(1) (1998) 127.

\bibitem{steane1999efficient}
A.~M. Steane, Efficient fault-tolerant quantum computing, Nature 399~(6732)
  (1999) 124--126.

\bibitem{bacon2006operator}
D.~Bacon, Operator quantum error-correcting subsystems for self-correcting
  quantum memories, Physical Review A 73~(1) (2006) 012340.

\bibitem{bravyi1998quantum}
S.~B. Bravyi, A.~Y. Kitaev, Quantum codes on a lattice with boundary, arXiv
  preprint quant-ph/9811052 (1998).

\bibitem{kitaev2003fault}
A.~Y. Kitaev, Fault-tolerant quantum computation by anyons, Annals of physics
  303~(1) (2003) 2--30.

\bibitem{knill1996concatenated}
E.~Knill, R.~Laflamme, Concatenated quantum codes, arXiv preprint
  quant-ph/9608012 (1996).

\bibitem{nielsen2010quantum}
M.~A. Nielsen, I.~L. Chuang, Quantum computation and quantum information,
  Cambridge university press, 2010.

\bibitem{gottesman2010introduction}
D.~Gottesman, An introduction to quantum error correction and fault-tolerant
  quantum computation, in: Quantum information science and its contributions to
  mathematics, Proceedings of Symposia in Applied Mathematics, Vol.~68, 2010,
  pp. 13--58.

\bibitem{aharonov1997fault}
D.~Aharonov, M.~Ben-Or, Fault-tolerant quantum computation with constant error,
  in: Proceedings of the twenty-ninth annual ACM symposium on Theory of
  computing, 1997, pp. 176--188.

\bibitem{aliferis2007accuracy}
P.~Aliferis, D.~Gottesman, J.~Preskill, Accuracy threshold for postselected
  quantum computation, arXiv preprint quant-ph/0703264 (2007).

\bibitem{knill1998resilient}
E.~Knill, R.~Laflamme, W.~H. Zurek, Resilient quantum computation: error models
  and thresholds, Proceedings of the Royal Society of London. Series A:
  Mathematical, Physical and Engineering Sciences 454~(1969) (1998) 365--384.

\bibitem{knill1996threshold}
E.~Knill, R.~Laflamme, W.~Zurek, Threshold accuracy for quantum computation,
  arXiv preprint quant-ph/9610011 (1996).

\bibitem{zalka1996threshold}
C.~Zalka, Threshold estimate for fault tolerant quantum computation, arXiv
  preprint quant-ph/9612028 (1996).

\bibitem{aliferis2005quantum}
P.~Aliferis, D.~Gottesman, J.~Preskill, Quantum accuracy threshold for
  concatenated distance-3 codes, arXiv preprint quant-ph/0504218 (2005).

\bibitem{chao2018quantum}
R.~Chao, B.~W. Reichardt, Quantum error correction with only two extra qubits,
  Physical review letters 121~(5) (2018) 050502.

\bibitem{chamberland2018flag}
C.~Chamberland, M.~E. Beverland, Flag fault-tolerant error correction with
  arbitrary distance codes, Quantum 2 (2018) 53.

\bibitem{tansuwannont2020flag}
T.~Tansuwannont, C.~Chamberland, D.~Leung, Flag fault-tolerant error
  correction, measurement, and quantum computation for cyclic
  calderbank-shor-steane codes, Physical Review A 101~(1) (2020) 012342.

\bibitem{chao2018fault}
R.~Chao, B.~W. Reichardt, Fault-tolerant quantum computation with few qubits,
  NPJ Quantum Information 4~(1) (2018) 42.

\bibitem{chamberland2019fault}
C.~Chamberland, A.~W. Cross, Fault-tolerant magic state preparation with flag
  qubits, Quantum 3 (2019) 143.

\bibitem{reichardt2020fault}
B.~W. Reichardt, Fault-tolerant quantum error correction for steane’s
  seven-qubit color code with few or no extra qubits, Quantum Science and
  Technology 6~(1) (2020) 015007.

\bibitem{quan2022implementation}
D.~Quan, C.~Liu, X.~Lv, C.~Pei, Implementation of fault-tolerant encoding
  circuit based on stabilizer implementation and “flag” bits in steane
  code, Entropy 24~(8) (2022) 1107.

\bibitem{portmann2022security}
C.~Portmann, R.~Renner, Security in quantum cryptography, Reviews of Modern
  Physics 94~(2) (2022) 025008.

\bibitem{steane1997active}
A.~M. Steane, Active stabilization, quantum computation, and quantum state
  synthesis, Physical Review Letters 78~(11) (1997) 2252.

\bibitem{knill2005quantum}
E.~Knill, Quantum computing with realistically noisy devices, Nature 434~(7029)
  (2005) 39--44.

\bibitem{yang2007upper}
L.~Yang, Y.~Chen, An upper bound to the number of gates on single qubit within
  one error-correction period of quantum computation, arXiv preprint
  arXiv:0712.3197 (2007).

\bibitem{yang2008decoherent}
L.~Yang, Y.~Chen, A decoherent limit of fault-tolerant quantum computation
  driven by coherent fields, in: Quantum Optics, Optical Data Storage, and
  Advanced Microlithography, Vol. 6827, SPIE, 2008, pp. 49--54.

\bibitem{yang2013full}
L.~Yang, B.~Yang, Y.~Chen, Full quantum treatment of rabi oscillation driven by
  a pulse train and its application in ion-trap quantum computation, IEEE
  Journal of Quantum Electronics 49~(8) (2013) 641--651.

\bibitem{yang2020effect}
B.~Yang, L.~Yang, Effect on ion-trap quantum computers from the quantum nature
  of the driving field, Science China Information Sciences 63 (2020) 1--15.

\bibitem{yang2013post}
L.~Yang, R.-R. Zhou, On the post-quantum security of encrypted key exchange
  protocols, arXiv preprint arXiv:1305.5640 (2013).

\end{thebibliography}

\end{document}